\documentclass[%
 aip,
 amsmath,amssymb,
 reprint,%
]{revtex4-2}

\usepackage{natbib}
\usepackage{graphicx}
\usepackage{dcolumn}
\usepackage{bm}

\usepackage[utf8]{inputenc}
\usepackage[T1]{fontenc}
\usepackage{mathptmx}
\usepackage{caption}

\begin{document}

\preprint{AIP/123-QED}

\title{Artificial neural network-based reduced-order modeling for turbulent wake of a finite wall-mounted square cylinder}

\author{Mustafa Z. Yousif}
\affiliation{School of Mechanical Engineering, Pusan National University, 2, Busandaehak-ro 63beon-gil, Geumjeong-gu, Busan, 46241, Rep. of KOREA}

\author{Hee Chang Lim}
\email[]{Corresponding author, hclim@pusan.ac.kr}
\thanks{}
\affiliation{School of Mechanical Engineering, Pusan National University, 2, Busandaehak-ro 63beon-gil, Geumjeong-gu, Busan, 46241, Rep. of KOREA}

\date{\today}

\begin{abstract}
This study presents an artificial neural network and proper orthogonal decomposition (POD)-based reduced-order model (ROM) of turbulent flow around a finite wall-mounted square cylinder. The proposed model is suitable for turbulent wake control applications because it can predict the dynamics of the main features of the flow field without computing Navier-Stokes equations. Long short-term memory neural network (LSTM NN) and bidirectional long short-term memory neural network (BLSTM NN) are used to predict the temporal evolution of the POD time coefficients at different planes along the height of the obstacle. The improved delayed detached-eddy simulation (IDDES) is performed to generate the training datasets. Transfer learning (TL) approach is utilized  in the training process by using the weights of the LSTM/BLSTM NN that are used to predict the POD time coefficients of the planes at lower elevations to initialize the weights of the networks at higher elevations along the height of the obstacle. The use of TL results in a remarkable improvement in the capability of the LSTM/BLSTM NN prediction compared with the one when the network is initialized with random weights. BLSTM NN shows better results compared with LSTM NN in terms of training and prediction error, indicating that the BLSTM-POD model is more suitable to be used as a ROM for predicting the turbulent wake. Furthermore, the temporal behavior of the time coefficients is carefully examined using the phase space plots and Poincar$\acute{e}$ sections. The results of using different lengths of the prediction time window showed that the prediction error of the POD time coefficients increases as the prediction time window increases and the error increasing rate decreases with the ranking of the POD time coefficients.
\end{abstract}

\maketitle

\section{Introduction}\label{sec:introduction}
The study of wake dynamics behind finite wall-mounted cylinders (FWMCs) has received considerable attention over the past few decades. From skyscrapers, chimneys, and submarine appendages to electronic chips, many engineering structures can be approximated as FWMCs. The study of the wakes behind the finite wall-mounted obstacles helps understand and control the flows around the engineering structures. However, this type of flows has a complex behavior as it is three-dimensional, this is different from the the case of infinite-length cylinder where the flow is mostly two-dimensional \cite{Bourgeoisetal2011, Park&Lee2000, Sattarietal2012}. For the FWMC, the flow is resulted from the effect of the downwash from the free-end, the von K$\acute{a}$rm$\acute{a}$n spanwise vortex shedding, and the base upwash\cite{Saeedietal2014, Saha2013}. \par 

This complex nature of the flow requires an appropriate experimental setup and a numerical model that can accurately capture most of the flow-field features and provide high-fidelity data. The recent advances in experimental and computational techniques have helped in visualizing and analyzing the flow structure around the obstacles and helped obtain enormous amounts of data for instance, from practical image velocimetry (PIV) and direct numerical simulation (DNS) studies. Nevertheless, understanding the main features of the flow field based on such big data is difficult due to the complexity of the flow behavior as well as the existence of redundant data. Furthermore, building an optimum model that has a low computational cost is an important factor in turbulent flow-control applications \cite{Tangetal1996}. For the finite wall-mounted obstacles, building an analytical low-order model that can describe the shedding process is often not applicable because the shedding process changes with the obstacle length and the flow becomes three-dimensional as it approaches the free-end due to the downwash effect \cite{Wang&Zhou2009, Yousif&Lim2020}. \par

Data-driven methods such as proper orthogonal decomposition (POD) \cite{Lumley1967} and dynamic mode decomposition (DMD) \cite{Schmid2010} have been widely used in fluid dynamics for ROM. In these methods, the flow fields are mapped  to a low-dimensional space that is represented by an optimal set of modes. In POD, these modes are based on the energy content of the flow, while in DMD, they are related to the associated frequencies. The traditional way of modeling the dynamics of the flow in these methods is to perform the Galerkin projection approach \cite{Burkardtetal2006, Carlbergetal2011, Rapun&Vega2010}. The Galerkin projection method uses the spatio-temporal dynamics that are captured from POD or DMD to build a low-order model that can evolve over time instead of the full-order Navier-Stokes equations. This can result in a cheaper computation because the low-order model contains considerably fewer degrees of freedom \cite{Rowleyetal2000}. However, the Galerkin projection approach exhibits an unstable behavior, even for simple canonical cases \cite{Akhtaretal2009, Rempfer2000}. Furthermore, this approach still requires effort in terms of the computational model setup. These issues led to the development of other methods \cite{Shindeetal2016}. However, for complex turbulent flow problems, an accurate approach is required to capture the non-linear dynamics of the flow.\par

machine learning learning has shown considerable promise in modeling complex and non-linear systems in diverse fields like image processing, robotics, weather forecasting, etc. In recent years, machine learning has been used extensively in fluid dynamics for experimental data processing, turbulence closure modeling, data-driven control as well as reduced-order modeling \cite{Bruntonetal2020, Kutz2017}. For reduced-order modeling, promising results have been obtained for the prediction of turbulent flows using various machine learning algorithms that based on artificial neural networks such as multi-layer perceptrons (MLPs) (also known as fully-connected-layer neural networks), convolution neural networks (CNNs), and recurrent neural networks (RNNs) including the long-short-term memory neural networks (LSTM NNs).\par

It has been established that the results obtained from a fully connected-layer neural network with a single hidden layer and linear activation function are equivalent to that obtained from a POD \cite{Baldi&Hornik1989, Bourlard&Kamp1988}. This shows that the neural networks are based on data-driven training and can be compared and used with other traditional data-driven methods. Milano and Koumoutsakos \cite{Milano&Koumoutsakos2002} showed that by using POD-based neural networks with non-linear activation functions, the prediction of the near-wall velocity field is improved significantly, indicating better predictive capabilities of the neural networks. \par

G$\ddot{u}$emes {\it et al.} \cite{Guemesetal2019} used CNN to predict the POD time coefficients from streamwise wall-shear stress measurements that were used to build a reduced-order reconstruction of the velocity fields. They recorded that CNN significantly outperformed the extended POD (EPOD), which is considered a linear method. Murata {\it et al.} \cite{Murataetal2020} used a CNN auto-encoder and LSTM NN to obtain a low-order representation of flow around bluff bodies of various shapes. Their instantaneous results, as well as statistical calculations, showed that the flow fields conformed to be well reproduced. Motivated by the capabilities of LSTM NNs in modeling sequential data, Z. Wang {\it et al.} \cite{Wangetal2018} used LSTM NN to build a POD-based reduced-order model of an ocean gyre and flow past a cylinder by modeling the POD time coefficients. They obtained a significant reduction in the computational cost with reasonable predictive accuracy. Mohan and Gaitonde \cite{Mohan&Gaitonde2002} applied LSTM and bidirectional long-short term (BLSTM) NNs to predict the POD time coefficients that were used to build a reduced-order model of canonical DNS datasets. In their study, LSTM NN outperformed BLSTM NN, suggesting that BLSTM NN is more suitable for the flow with periodic or quasi-periodic temporal behavior such as two-dimensional bluff body wake shedding.\par

In this study, a deep neural network-based approach to build a reduced-order model using LSTM and BLSTM NNs combined with POD is applied to datasets obtained from an improved delayed detached-eddy simulation (IDDES) of flow past a finite wall-mounted square cylinder with an aspect ratio of four at Reynolds number $Re_D=12 000$. The objective of this approach is to investigate the applicability of the LSTM-POD and BLSTM-POD models for predicting the instantaneous coherent structure of the flow field along the height of the obstacle. \par

The remainder of this paper is organized as follows: In Section II, the case geometry and the numerical model used to generate the training data are explained, and the simulation results are validated with the available experimental and DNS data. POD is briefly explained in Section III. Section IV provides a brief introduction to LSTM and BLSTM NNs while Section V explains TL approach used in this study. The methodology of building LSTM-POD/BLSTM-POD ROM is explained in Section VI. The results of this study are thoroughly discussed in Section VII. Finally, the conclusions of this study are presented in Section VIII. \par

\section{Preparing the training data}

\subsection{Case geometry and boundary conditions}

\begin{figure*}
\centering 
\includegraphics[angle=0, trim=0 0 0 0, width=0.8\textwidth]{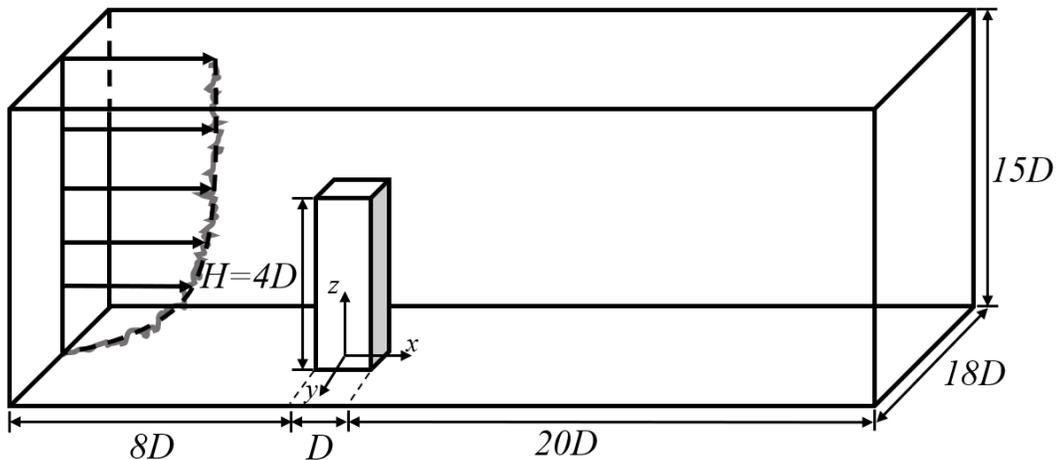}
\caption[]{Schematic of the computational domain.} 
\label{fig:1_domain_diagram}
\end{figure*}

The choice of a finite-wall mounted square cylinder with an aspect ratio of four in this study is because it is well studied experimentally \cite{Bourgeoisetal2011} and numerically \cite{Saeedietal2014}. Furthermore, the sharp leading edges of the square cylinder ensure that the shedding process is not affected by the change in $Re_D$, at least over the range tested by Sattari {\it et al.} \cite{Sattarietal2012}. The schematic diagram of the numerical domain for the finite wall-mounted square cylinder is shown in Fig.~\ref{fig:1_domain_diagram}.

The square cylinder has a height $H = 4D$, where $D$ is its width, and $Re_D = 12,000$ ( based on the cylinder width and free stream velocity $U_\infty$). The size of the domain is equal to $29D \times 18D \times 15D$. Therefore, the blockage ratio is found to be approximately 1.5\%, which is lower than the maximum value of 3\% recommended by Franke et al. (Franke et al., 2004). The obstacle is located at a distance of $8D$ downstream from the inlet and $20D$ upstream from the outlet. The distance of the obstacle from each side of the domain is set to $8.5D$ while the distance from the end of the obstacle to the top surface of the domain is set to $11D$. In turbulent flows simulations, applying turbulent inflow conditions is crucial for  the simulation reliability. Thus,  the vortex method \cite{Matheyetal2006, Sergent2002} is implemented in this study to generate unsteady turbulent inflow conditions. Pressure outlet boundary condition is assigned for the outlet plane of the domain, and periodic boundary condition is applied to the spanwise direction. No-slip condition is applied to the obstacle walls and the ground, while a symmetry plane is applied to the top surface of the computational domain. \par

\subsection{Governing equations and turbulence modeling}

The momentum equation for an incompressible viscous fluid is:

\begin{equation}\label{EQ1}
\left(  \frac{\partial \bf{u}}{\partial t} + \bf{u} \cdot \nabla \bf{u} \right) = - \frac{1}{\rho} \nabla p + \nu \nabla^2 \bf{u},
\end{equation}

\noindent where ${\bf u}$ is the velocity of the fluid, $\rho$ is the density, $p$ is the pressure, and $\nu$ is the kinematic viscosity. The incompressibility of the fluid is expressed by the continuity equation: \par

\begin{equation}\label{EQ2}
\nabla \cdot \bf{u} = 0.
\end{equation}

Different numerical approaches have been used to solve and model the flow parameters in Eqs. (\ref{EQ1}) and (\ref{EQ2}). DNS resolves all scales of turbulent flow structures, from the large energy-containing scales to the smallest Kolmogorov length scale, in the dissipation range \cite{Kolmogorov1941}. To ensure this, an extremely fine grid size is required that results in a high computational cost, thus making DNS impractical for several applications, especially those with a high $Re$ value and complex geometries \cite{Moin&Mahesh1998}. The Reynolds-Averaged Navier-Stokes (RANS) equations solve only the time-averaged flow field, i.e., the effects of the fluctuating components are considered through the modeled Reynolds stresses. Thus, the RANS model is not suitable to describe the details of the flow field such as the instantaneous effect of multi-scale eddies \cite{Schmid2010, Wilcox1993}. \par

In LES, Kolmogorov's first hypothesis is considered, i.e., the small eddies in the turbulent flow tend to exhibit universal behavior. This property allows the modeling of small eddies and solving for the large eddies \cite{Smagorinsky1963, Stolzetal2001}. This process can be performed using spatial filters to separate the large scales from the small ones. The small-scale structures (small eddies) are modeled and introduced as a subgrid-scale (SGS) term in the filtered Navier–Stokes equations. Thus, LES is considerably more accurate as compared to RANS for describing the flow characteristics. Nevertheless, it requires high computational effort, which makes it impractical to simulate flow around engineering structures at high $Re$ values \cite{Wangetal2019}. Hybrid RANS-LES models, such as the DES, have been introduced by Spalart and co-authors \cite{Spalartetal1997, Spalart2000, Travinetal2000} to bridge the gap between LES and RANS. DES has been designed to simulate the flow at a high $Re$ value and to flow around bodies with massive flow separation. \par

By applying RANS near the solid walls within the attached boundary layer and using LES outside the boundary layer in the separated flow region, the grid size for numerical simulations can be reduced significantly in the near-wall region, compared to that required by LES. DES is a promising turbulence modeling tool that can be applied for simulations including wall-bounded flows at high $Re$ values. However, some undesirable phenomena such as modeled stress depletion and grid-induced separation may occur and affect the accuracy of the model \cite{Spalart2009}. This could be attributed to the mesh resolution inside the boundary layer that may result in DES attempting LES rather than RANS in the near-wall region. To overcome these limitations of DES, variant versions including delayed DES (DDES) \cite{Spalartetal2006} and IDDES \cite{Shuretal2008} have been introduced that are suitable for different grid spacing inside the boundary layer regardless of the boundary layer thickness. Therefore, considering the relatively high Reynolds number used in this study and the complex flow that is expected to be obtained from the simulation, the Spalart–Allmaras IDDES (S-A IDDES) is used in this study to simulate the flow.

\subsection{Numerical framework}

The computational fluid dynamics (CFD) finite-volume code OpenFOAM-5.0x is used to perform the simulation. A structured hexahedral mesh is used for spatial discretization. Local mesh refinement is applied using the stretching mesh technique in regions where high-velocity gradients are expected. Grid-independence tests are performed using three different grid sizes: $132 \times 122 \times 90$, $152 \times 142 \times 110$, and $172 \times 162 \times 130$. The overall time-averaged drag coefficient and Strouhal number are chosen as the parameters for the test. As shown in Table~\ref{tab:gridtest}, less than 1\% variation is observed in the drag coefficient while there is no variation in Strouhal number with the progressive refinement of the mesh, ultimately converging with the grid-independence test. Considering the maximum dimensionless first cell spacing $y^+$, the finest mesh size of $172 \times 162 \times 130$ is chosen for this study. \par

\begin{table}
  \begin{center}
  \begin{tabular}{c  c  c  c}\hline\hline \\
  No of grid elements      &  $N_x \times N_y \times N_z $ \footnote[1]{$N_x , N_y$ , and $N_z $ represent the number of grid points distributed on the walls of the obstacale.}  & Overall drag, $C_D$  & Strouhal No., $St$ \\[3pt] \hline \\
  $132 \times 122 \times 90 $   & $22 \times 22 \times 40 $   &  1.212  & 0.101      \\  
  $152 \times 142 \times 110 $  & $22 \times 22 \times 50 $  &  1.238  & 0.104 \\
  $172 \times 162 \times 130 $  & $22 \times 22 \times 60 $  &  1.241  & 0.104 \\ \\ \hline \hline
  \end{tabular}
  \caption{Grid independence test.}
  \label{tab:gridtest}
  \end{center}
\end{table}

\begin{figure*}
\centering 
\includegraphics[angle=0, trim=0 0 0 0, width=0.8\textwidth]{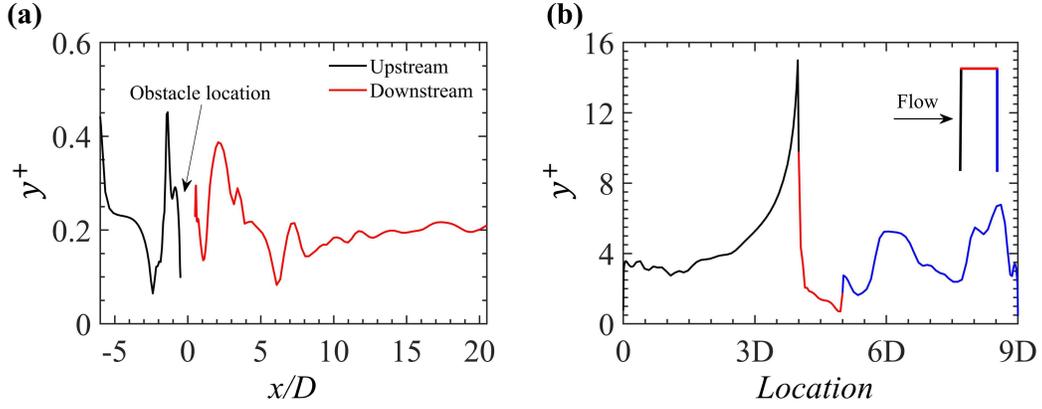}
\caption[]{Distribution of $y^+$ in the central $x-z$ plane ($y/D = 0$): (a) along the ground surface, and (b) around the obstacle surfaces.}
\label{fig:2_y_plus}
\end{figure*}

Figure~\ref{fig:2_y_plus}  shows the distribution of $y^+$ along the ground surfaces and around the obstacle surfaces in the central $x-z$ plane ($y/D = 0$). The maximum value of $y^+$ is less than 0.5 on the ground while it is approximately 15 on the obstacle. These values of $y^+$ are within the applicable range for IDDES \cite{Spalart2009}. Advancement is done through a time-step of $6 \times 10^{-6}$s while maintaining the mean Courant number (CFL) at 0.52 and the maximum CFL at 0.95 to ensure a stable simulation. The approximate time required for every vortex shedding cycle is $8.5 \times 10^{-3}$s. To ensure that the flow reached a statistically stationary state, the simulation is first extended over 30 shedding cycles. Subsequently, the flow statistics are collected through approximately 60 shedding cycles. The PIMPLE algorithm, which is a combination of the PISO algorithm (pressure implicit split operator) and the SIMPLE algorithm (semi-implicit method for pressure-linked equations), is employed to solve the coupled pressure momentum system. The convective fluxes are discretized with a second-order-accurate linear upwind scheme, and all other discretization schemes used in the simulation have second-order accuracy. To perform parallel computing, the computational domain is divided into 80 sub-domains and  Scotch method is used to decompose the domains. Correspondingly, 80 processors were used to solve the flow field at each time-step. All the computations are performed using a computer cluster with Intel Xeon Skylake (Gold 6148) 2.4 GHz processor. 

\subsection{Validation of the numerical results}

\begin{figure*}
\centering 
\includegraphics[angle=0, trim=0 0 0 0, width=0.8\textwidth]{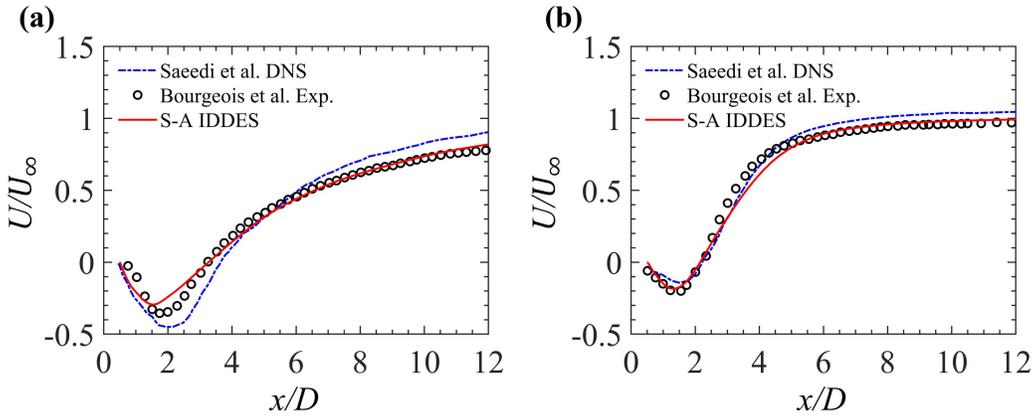}
\caption[]{Streamwise profile of the mean streamwise velocity at two different elevations in the central $x-z$ plane ($y/D = 0$): (a) $z/D = 1$, and (b) $z/D =3$.} 
\label{fig:3_v_x_validation}
\end{figure*}

To ensure that the simulation produced reliable training data, the first and second-order flow statistics obtained from the simulation are compared with the available experimental measurements of Bourgeois {\it et al.} \cite{Bourgeoisetal2011} and the DNS data of Saeedi {\it et al} \cite{Saeedietal2014}. Figure~\ref{fig:3_v_x_validation} shows the recovery of the mean streamwise velocity, $U$ in the central $x-z$ plane ($y/D = 0$) for two different elevations behind the obstacle, $z/D =$ 1 and 3. For both elevations, the velocity has negative values near the wall due to the reverse flow as a result of recirculation. For $x/D \approx$ 3, the velocity value becomes positive and the flow recovers to the free stream velocity with a higher rate at elevation $z/D$ = 3. Both Figs.~\ref{fig:3_v_x_validation}(a) and (b) correlate with the experimental and DNS results. \par

\begin{figure*}
\centering 
\includegraphics[angle=0, trim=0 0 0 0, width=0.8\textwidth]{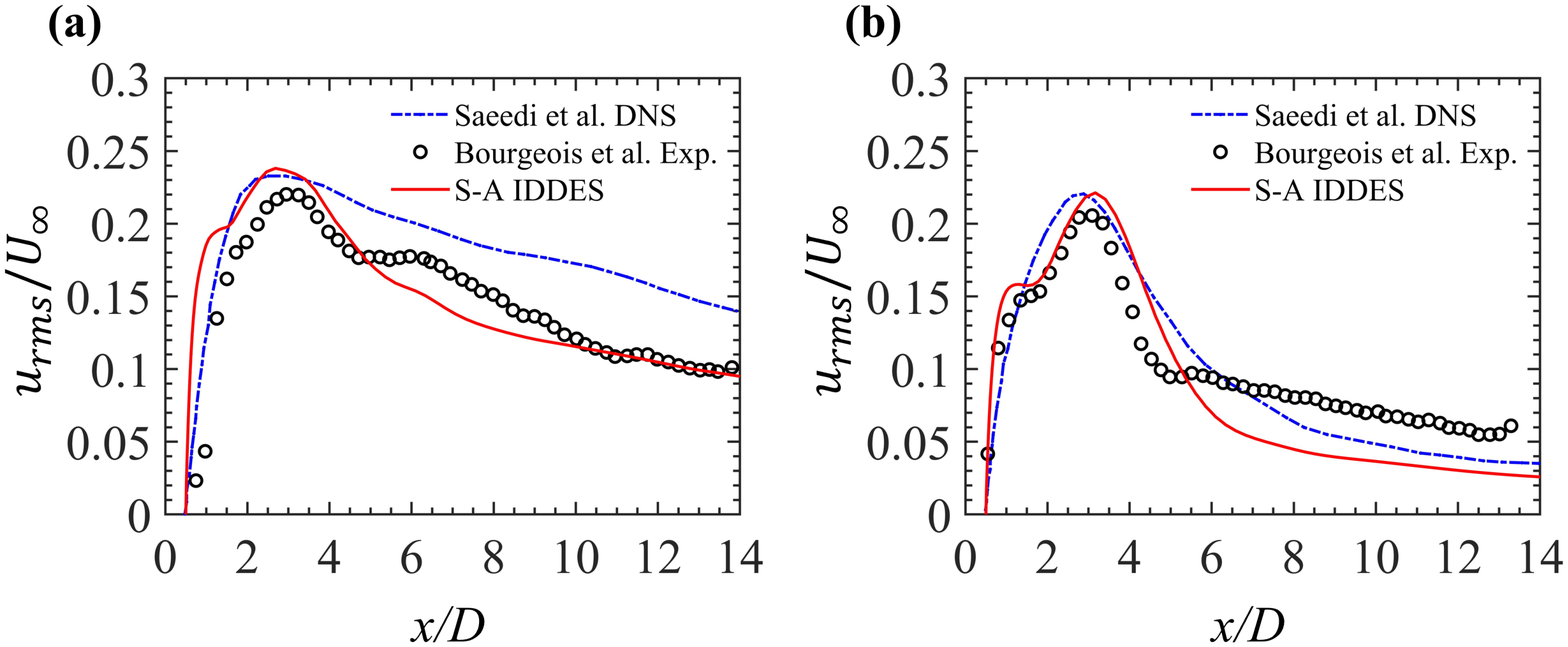}
\caption[]{Streamwise profile of the turbulence intensity at two different elevations in the central $x-z$ plane ($y/D = 0$): (a) $z/D = 1$, and (b) $z/D =3$.} 
\label{fig:4_vrms_validation}
\end{figure*}

Figure~\ref{fig:4_vrms_validation} shows the streamwise profile of the turbulence intensity, $u_{rms}$ obtained from the simulation, experimental measurements, and DNS data in the central $x-z$ plane ($y/D$ = 0) for two different elevations $z/D$ = 1 and 3. The values of turbulent intensity have been well reproduced at the near-wall peak region for both elevations. However, under-prediction of the values in the far downstream of the obstacle at an elevation $z/D = 3$ can be observed in Fig.~\ref{fig:4_vrms_validation}(b). This is attributed to the turbulent level attenuation in this region, which demands more computational effort. \par

\begin{figure*}
\centering 
\includegraphics[angle=0, trim=0 0 0 0, width=0.8\textwidth]{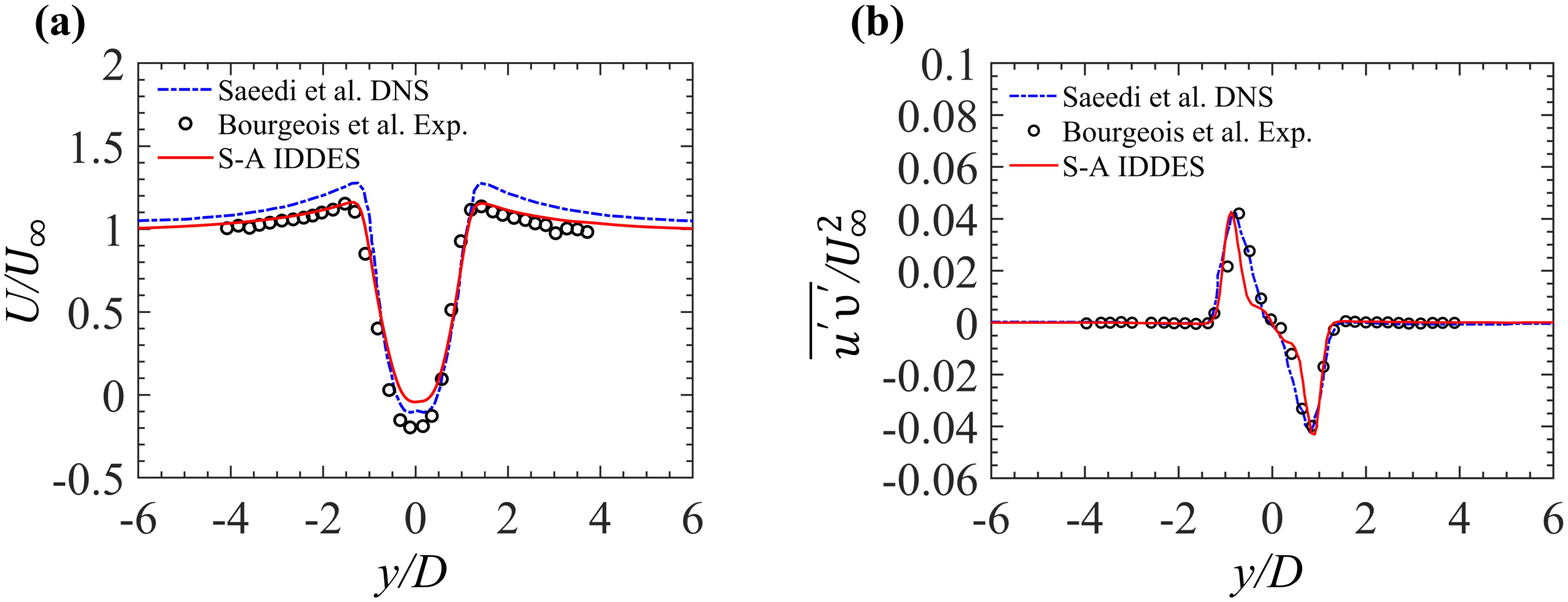}
\caption[]{Spanwise profile of (a) the mean streamwise velocity, and (b) mean Reynolds stress component $\overline{u'\upsilon'}$ at elevation $z/D = 3$ and streamwise location $x/D = 2$.} 
\label{fig:5_v_uv_validation}
\end{figure*}

The mean streamwise velocity profile along the lateral direction at elevation $z/D = 3$ and streamwise location, $x/D = 2$ is shown in Fig.~\ref{fig:5_v_uv_validation} (a). The streamwise velocity values become negative at $y/D$ between -0.5 and 0.5 indicating the effect of recirculation inside the wake. The velocity profile obtained from the numerical simulation complies with the experimental measurements and DNS results. However, the streamwise velocity values are slightly over-predicted in the recirculation region. Figure~\ref{fig:5_v_uv_validation} (b) compares the numerical results with the experimental measurements and DNS results of the mean Reynolds stress component $\overline{u'\upsilon'}$ ($\upsilon'$ is the fluctuating part of the spanwise velocity component) at an elevation $z/D = 3$ along the lateral direction and at the streamwise location $x/D = 2$. As shown in the figure, the mean Reynolds stress profile has maximum negative and positive values at $y/D = \pm 1$ that decrease rapidly in the lateral direction indicating that the shear stress accrues due to the combination of the vortices induced behind the obstacle. A favorable agreement with the experimental and DNS results can be observed from the figure. Here, the S-A IDDES model successfully reproduces the flow field around and behind the obstacle with commendable accuracy. This is attributed to the turbulent inflow conditions obtained from the vortex method and the small blockage ratio used in this study.

\section{Proper orthogonal decomposition}

POD \cite{Lumley1967} is widely used in fluid dynamics to extract the main features of the flow field, which is known in literature as the coherent structure \cite{Berkoozetal1993}. POD decomposes the fluctuating parts of the velocity components into spatial orthogonal modes $\psi_m$ and corresponding time coefficients $a_m (t)$. The extracted modes represent the most coherent structure of the flow field according to the energy content of the modes. In this study, POD was applied to different $x-y$ planes along the obstacle height. \par

\begin{equation}\label{EQ3}
u' (x, y, t) = \Sigma_{m=1}^{M} a_m (t) \phi_m (x, y),
\end{equation}

\noindent where $u'$ here represents the fluctuating part of the streamwise velocity component and $M$ is the number of the POD modes. The snapshot POD method \cite{Sirovich1987} was used in this study, which is based on computing the auto-correlation from numerous snapshots of the instantaneous flow field. The temporal modes are the eigenvectors that can be obtained by solving the eigenvalue problem of the auto-correlation matrix,

\begin{equation}\label{EQ4}
C = V V^T,
\end{equation}

where $V$ is a matrix containing the fluctuating part of the velocity:

\begin{equation}\label{EQ5}
V = 
\begin {bmatrix}
{u'}_1^1 & \cdots & {u'}_1^N  \\
\vdots & \ddots & \vdots \\
{u'}_M^1 & \cdots &  {u'}_M^N \\
\end {bmatrix},
\end{equation}

\noindent where $N$ represents the position in each snapshot (instantaneous) data of the flow. The corresponding sets of $M$ eigenvalues $\lambda_m $ represent the contribution to the turbulent kinetic energy captured by the respective POD modes. The eigenvalues are arranged in decreasing order such that:

\begin{equation}\label{EQ6}
\lambda_1 \geq \lambda_2 \cdots \geq \lambda_M \geq 0.
\end{equation}

\section{Long short-term memory neural networks}

\begin{figure*}
\centering 
\includegraphics[angle=0, trim=0 0 0 0, width=0.5\textwidth]{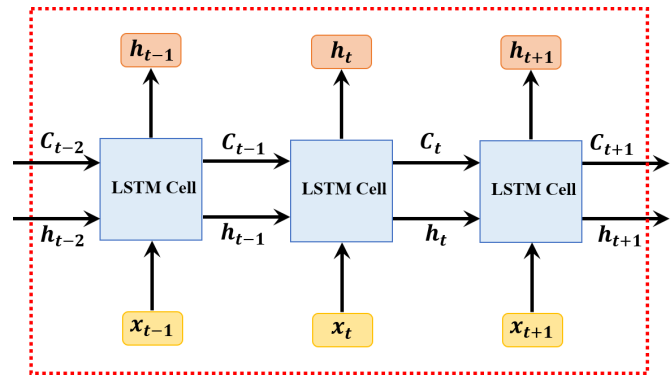}
\caption[]{LSTM NN Architecture.} 
\label{fig:6_lstm_1}
\end{figure*}

LSTM NN \cite{Hochreiter&Schmidhuber1997} has been widely used in different fields such as speech recognition, language translation, as well as time-series prediction. LSTM NN is a special variant of RNNs \cite{Rumelhartetal1986}, which is designed to overcome the stability issues and limitations of traditional RNNs such as blowing up or vanishing gradient \cite{Graves2012}. Because of its architecture, LSTM NN has a great ability to learn the temporal dependence from the data. These unique properties make LSTM NN suitable for time-series problems. The architecture of LSTM NN is shown in Fig.~\ref{fig:6_lstm_1}. The hidden layer in LSTM NN receives an input vector that contains the sequential data x and generates an output vector $h$, where $h$ represents the hidden state and $t$ is the step of the sequential data (a particular time-step of the time series data). The hidden layer maintains the hidden state with every iteration in the LSTM cell and updates it based on the cell input and the information from the previous time steps. This information is represented by the previous hidden state and the cell state ($C$). The cell state plays a crucial role in reducing the effect of short-term memory by carrying the relevant information throughout the processing of the sequence. 

\begin{figure*}
\centering 
\includegraphics[angle=0, trim=0 0 0 0, width=0.6\textwidth]{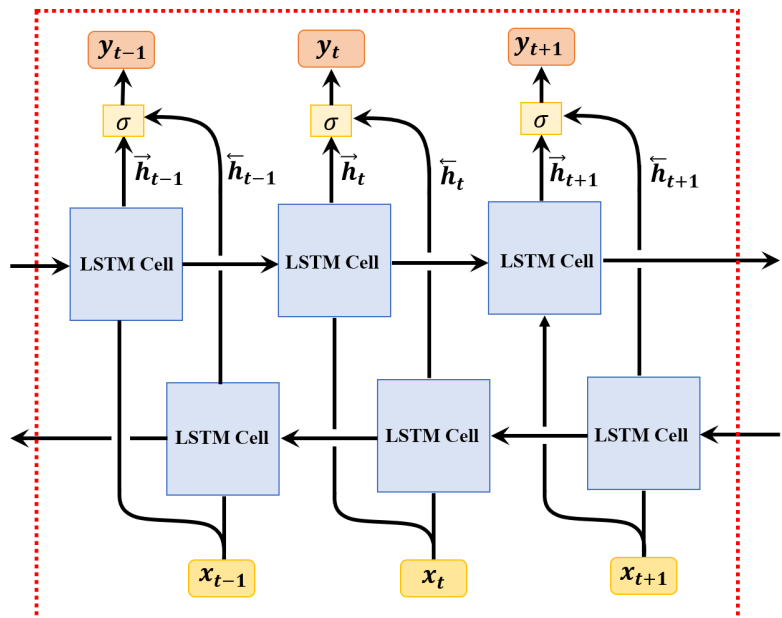}
\caption[]{BLSTM NN Architecture.} 
\label{fig:8_blstm}
\end{figure*}

Figure~\ref{fig:7_lstm_2} shows the basic architecture of the LSTM cell. LSTM cell contains three gates namely, input gate ($i$), output gate ($o$), and forget gate ($f$). The LSTM cell changes the flow of training information according to these gates. The input gate controls the flow of the input information into the LSTM cell. The forget gate filters the information from the previous time steps by deciding which information should be discarded or retained. The output gate controls the flow of the LSTM cell output. The equations that compute the output of the gates can be represented as follows:

\begin{figure*}
\centering 
\includegraphics[angle=0, trim=0 0 0 0, width=0.5\textwidth]{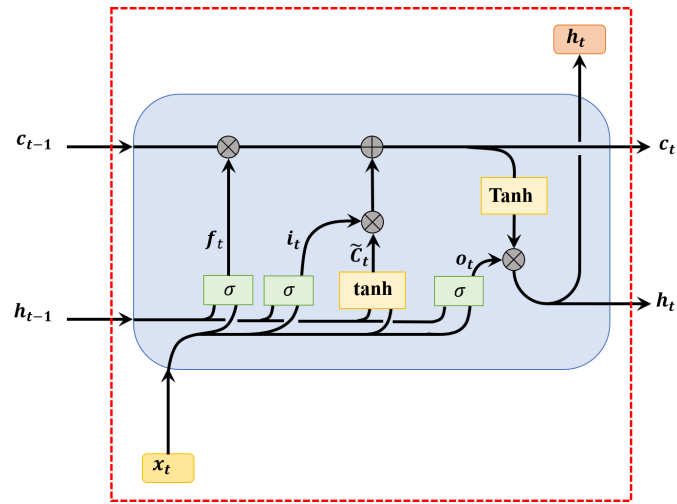}
\caption[]{LSTM cell Architecture.} 
\label{fig:7_lstm_2}
\end{figure*}

\begin{equation}\label{EQ7}
i_t = \sigma \left( W_i x_t + U_i h_{t-1} + b_i \right),
\end{equation}

\begin{equation}\label{EQ8}
f_t = \sigma \left( W_f x_t + U_f h_{t-1} + b_f \right),
\end{equation}

\begin{equation}\label{EQ9}
\tilde{C}_t = tanh\left( W_C x_t + U_C h_{t-1} + b_C \right),
\end{equation}

\begin{equation}\label{EQ10}
C_t = i_t \otimes \tilde{C}_t + f_t \otimes C_{t-1},
\end{equation}

\begin{equation}\label{EQ11}
o_t = \sigma \left( W_o x_t + U_o h_{t-1} + b_o \right),
\end{equation}

\begin{equation}\label{EQ12}
h_t = o_t \otimes tanh({C}_t),
\end{equation}

\noindent where $W_i$, $W_f$, $W_C$ and $W_o$  represent the weights that map the input to each of the gates and $U_i$, $U_f$, $U_C$ and $U_o$ are the weights related to the hidden state of the cell in the previous time step. Here, $\tilde{C}_t$  here represents the updated cell state. BLSTM NN \cite{Gravesetal2005, Graves&Schmidhuber2005} is a variant of LSTM NN that processes sequential data using two separate hidden layers in both forward and backward directions and connect both of them to the same output layer, as shown in Fig.~\ref{fig:8_blstm} . This architecture of BLSTM NN enables the two-way flow of the information so that the sequence can be trained in both forward and backward directions. BLSTM NN has shown an improved accuracy comparing with LSTM NN in speech recognition and language modeling \cite{Huangetal2015, Marchietal2014}. To compare the accuracy of these two networks, both LSTM and BLSTM NNs are used in this study to predict the time coefficients of POD modes for different $x-y$ planes along the height of the obstacle.

\section{Transfer learning}

The availability of sufficient training data and the time required for the training process are factors that should be considered for training neural networks. TL approach \cite{Weissetal2016} has been extensively used in different fields and showed promising results in terms of reducing the data and the time required for the training process. This method enables the reuse of pretrained network parameters to train another neural network that shares some features with the pretrained one even if it is designed for a different purpose than the pretrained network is designed for. Hence, TL helps to achieve optimum use of the available training data as well as the reduction of the training time compared with the network if it is trained from scratch, i.e., with randomly initialized parameters. In this study, TL is used to train LSTM/BLSTM NNs by transferring the weights from the networks that trained by using the data of the lower elevation flow field to the higher elevation one. LSTM/BLSTM NNs are trained using the data of the lower elevation flow field, $z/D = 0.02$ and used to initialize the weights in the training process of the network that predicts the POD time coefficients at the next $x-y$ plane with higher elevation. The weights are transferred to the network that is used for the subsequent elevation, and so on.

\section{Methodology}

\begin{figure*}
\centering 
\includegraphics[angle=0, trim=0 0 0 0, width=1.0\textwidth]{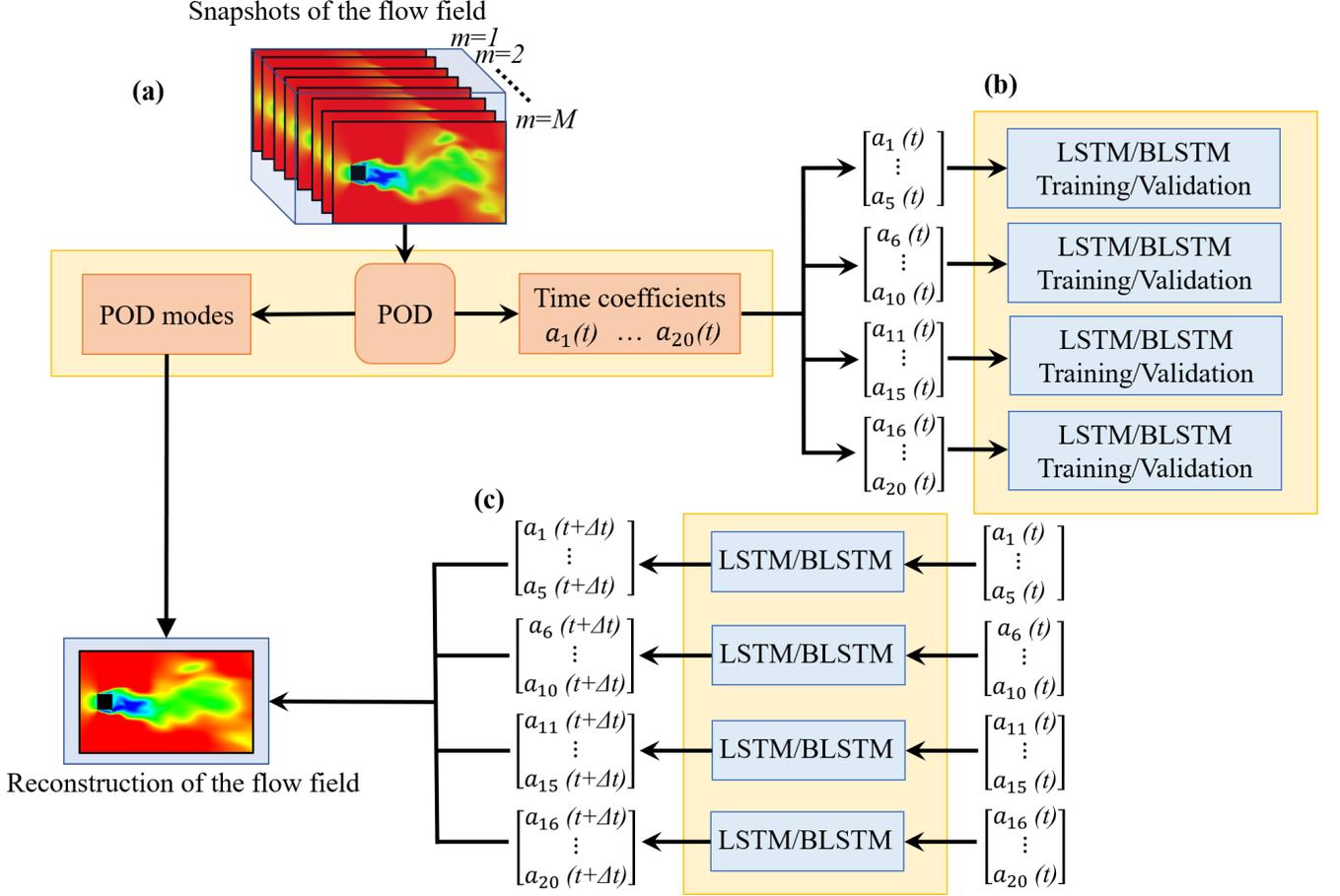}
\caption[]{Schematic of the LSTM/BLSTM-POD ROM.} 
\label{fig:9_procedure}
\end{figure*}

The main objective of this study is to build ROM based on POD modes and the temporal evolution of the time coefficients that can be obtained using LSTM or BLSTM NN. To provide a dataset that is adequate for performing the POD and training the LSTM and BLSTM NNs, four $x-y$ planes along the obstacle height are selected according to the expected behavior of the turbulent wake and 50 000 snapshots that are obtained from the IDDES, which are equivalent to 36 shedding cycles. The time coefficients obtained from the POD analysis are then used to train the LSTM and BLSTM NNs. The procedure that followed in this study is outlined in Fig.~\ref{fig:9_procedure}  and is expressed as follows,

\begin{enumerate}
\item POD analysis is performed on the datasets to extract the dominant POD modes and the corresponding time coefficients. The time coefficients are then used as datasets for LSTM and BLSTM NNs with 80\% of them as training data and 20\% as test data. Here the test datasets are used to validate the predicted time coefficients that are obtained from LSTM and BLSTM NNs.
\item LSTM and BLSTM NNs are trained using the training datasets and validated by comparing the predicted time coefficients of the future time-steps with the test datasets obtained from the POD analysis at the same time-steps.
\item The predicted flow field is then reconstructed using POD modes and the predicted time coefficients obtained from LSTM or BLSTM NN by using Eq. (\ref{EQ3}).\par
\end{enumerate}

The training datasets are normalized using the min-max normalization,

\begin{equation}\label{EQ13}
x_m (t) = \frac{a_m (t) - min (a_m (t) )}{max (a_m (t)) - min (a_m (t))},
\end{equation}

and decomposed into a series of short-time data samples and each sample is advanced by one time-step from the previous one. Each sample is used to predict the next time-steps that have the same time window length as the input sample. For instance, if a sample has a length of 10 time-steps, the model is trained to predict the time window with the length of the next 10 time-steps. \par

One LSTM/BLSTM layer with 500 units is used with one fully connected layer that has the number of units equal to the number of time coefficients in the prediction time window to predict the final output of the NN. As mentioned before, the TL approach is used to train the LSTM/BLSTM NN using the weights of the network that are used to predict the time coefficients at the previous $x-y$ plane starting from the lower elevation, i.e., at $z/D = 0.02$. Here, the learning rates for all the training processes are reduced by a factor of 20\% to prevent the optimizer from diverging too quickly from the weights that transferred from the previous LSTM/BLSTM NN. Four LSTM/BLSTM-POD models are used for every $x-y$ plane. Each model is trained with five sets of time coefficients, so the total number of time coefficients sets is twenty, which corresponds to the first twenty POD modes. All the weights are updated by back-propagation through the iteration time. 1,000 epochs are used for the training process and the adaptive moment estimation (Adam) optimizer \cite{Kingma&Ba2014} is used for both LSTM and BLSTM NNs.

\section{Results and discussion}

\subsection{POD analysis}

\begin{figure*}
\centering 
\includegraphics[angle=0, trim=0 0 0 0, width=1.0\textwidth]{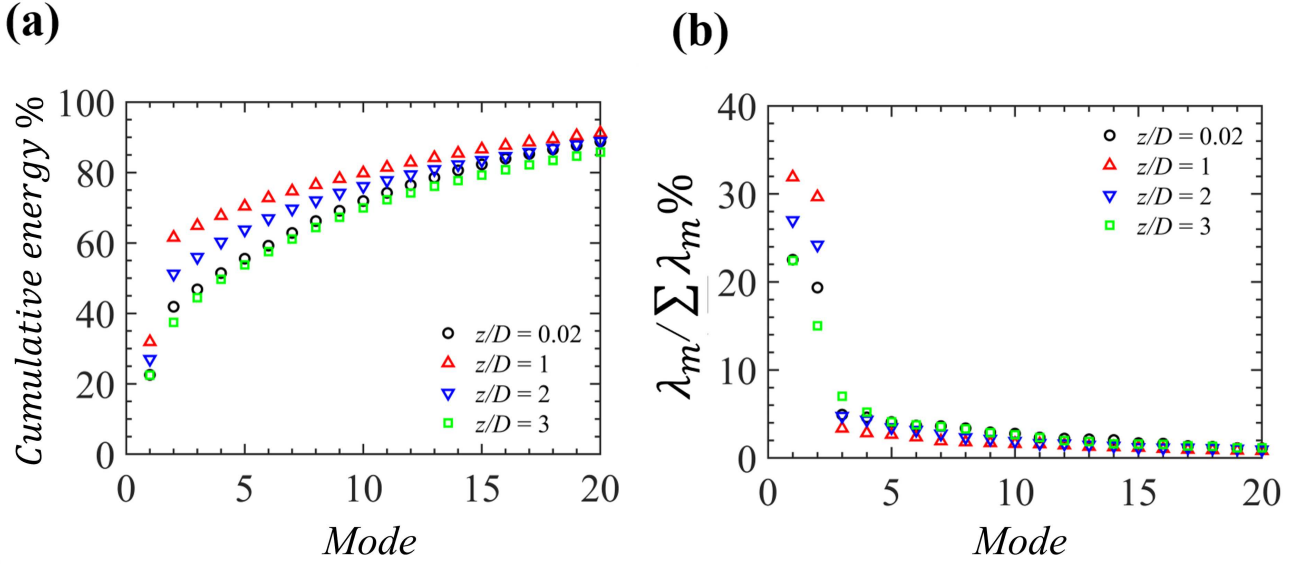}
\caption[]{Cumulative energy percentage of the eigenvalues (a), and the normalized eigenvalues of the first 20 POD modes (b).} 
\label{fig:10_modes_energy}
\end{figure*}

POD analysis was applied to four different $x-y$ planes along the height of the obstacle, $z/D$ = 0.02, 1, 2, and 3 respectively. A total of 50,000 snapshots across $6 \times 10^{-6}$s are used to perform POD at each elevation. Here, the same small time-steps used in the simulation are used in the POD analysis to capture most of the flow-field details that can enable better prediction. Figure~\ref{fig:10_modes_energy} shows the cumulative energy percentage of the eigenvalues and the normalized eigenvalues of the first 20 POD modes at different elevations along the obstacle length. Here, the eigenvalues represent the contribution of the corresponding POD modes to the total turbulent kinetic energy (TKE). As shown in Fig.~\ref{fig:10_modes_energy}(a), the maximum cumulative energy obtained from the first 20 modes is at $z/D$ = 1 (where the flow shows a quasi-periodic behavior) that is approximately 92\% of the total turbulent kinetic energy. However, as the periodic behavior becomes less at higher elevations and near the ground, the cumulative energy becomes less and reaches about 85\% at $z/D$ = 3. Similarly, in Fig.~\ref{fig:10_modes_energy}(b), the fractional contribution to the turbulent kinetic energy of the first two modes reaches 62\% at $z/D$ = 1 and 38\% at $z/D$ = 3. The normalized eigenvalues decrease smoothly for the rest of the modes with much lower values compared to the eigenvalues that corresponded to the first two modes. \par

\begin{figure*}
\centering 
\includegraphics[angle=0, trim=0 0 0 0, width=0.7\textwidth]{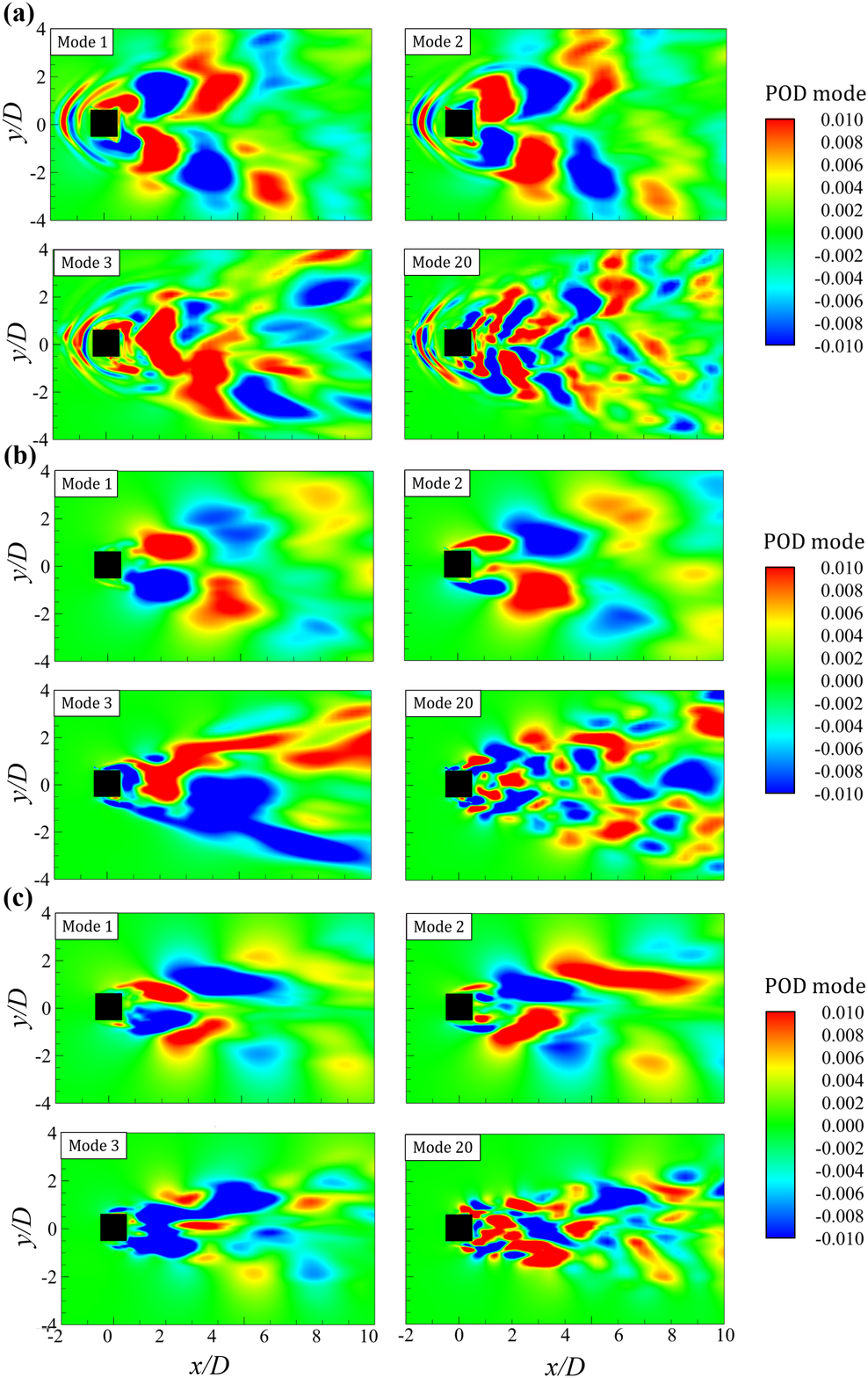}
\caption[]{Contour plots for different POD modes of the streamwise velocity field at three different elevations: (a) $z/D = 0.02$, (b) $z/D = 1$, and (c) $z/D = 3$.} 
\label{fig:11_modes}
\end{figure*}

Figure~\ref{fig:11_modes} shows the contour plots for different POD modes of the streamwise velocity field at three different elevations. The contour plot of the POD modes at $z/D = 2$ is not shown here because the behavior is the same as at $z/D = 1$. It can be seen clearly that from all the elevations along the obstacle height, the first and second modes contain the most coherent large-scale structure of the flow with clear alternative shedding behavior. However, the other modes exhibit less periodic behavior with an increase in the random behavior with the obstacle height. This change in behavior of the modes is attributed to the downwash effect from the free end of the obstacle, which tends to weaken the alternative shedding \cite{Yousif&Lim2020}. \par

\begin{figure*}
\centering 
\includegraphics[angle=0, trim=0 0 0 0, width=0.8\textwidth]{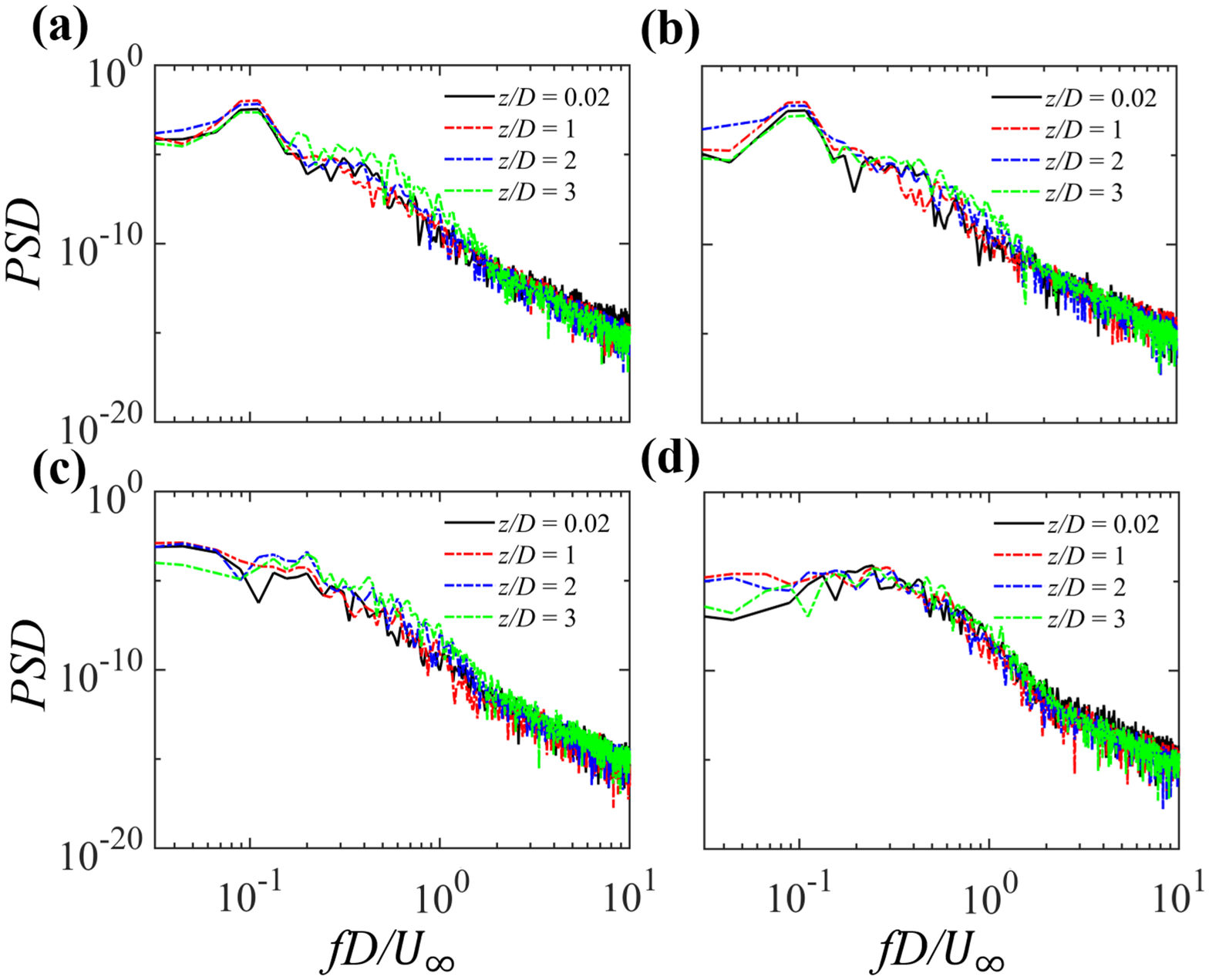}
\caption[]{Power spectrum density of different POD time coefficients obtained from the POD analysis at different elevations: (a) $a(t)_1$, (b) $a(t)_2$, and (c) $a(t)_3$, (d) $a(t)_{20}$.} 
\label{fig:12_modes_psd}
\end{figure*}

Figure~\ref{fig:12_modes_psd} shows the power spectrum density of different POD time coefficients obtained from the POD analysis at different elevations along the obstacle height. It can be seen from the figure that the time coefficients with the same POD mode ranking have similar frequency content along the obstacle height. This encourages us to perform TL according to the height of the obstacle with the aim that the time coefficients could share some similar features that can be transferred through the LSTM/BLSTM NN weights.

\subsection{TL and training error}

In this section, the effect of TL on the training accuracy and the capabilities of LSTM and BLSTM NNs are investigated. The LSTM/BLSTM NN is trained using samples with different lengths. We consider for the illustration here, a sample with a length of 10 time-steps. Table II shows the root-mean-square error of the LSTM/BLSTM NN output. The root-mean-square error is defined as:

\begin{equation}\label{EQ14}
\text{RMSE} = \sqrt{\frac{\Sigma_{i=1}^{k} (\hat{y}_i - y_i )^2}{k}},
\end{equation}

\noindent where $\hat{y}$ is the predicted value and $y$ is the actual value of the normalized time coefficient obtained from the test dataset. $k$ is the number of the time-steps. 

\begin{table*}
  \begin{center}
  \begin{tabular}{c  c  c  c c c}\hline\hline \\
  ~~~~~~& POD mode & $z/D = 0.02 $ &  $z/d = 1$ & $z/D = 2$ &$ z/D = 3$ \\[3pt] \hline \\
 LSTM without& 1& 0.0628 & 0.0599   &  0.0758  & 0.0666     \\  
 TL&2& 0.0948 & 0.0693  &  0.0916  & 0.0721 \\
 ~~~~~&3& 0.0963 &0.0825 &  0.1162 & 0.0802   \\ 
 ~~~~~&20& 0.1285 &0.1100 &  0.1731 & 0.1467    \\ \\

LSTM with& 1&  &0.0468   &  0.0532  & 0.0562     \\  
 TL&2& & 0.0562  &  0.0868  & 0.0713 \\
 ~~~~~&3&  &0.0588&  0.1051 & 0.0795   \\ 
 ~~~~~&20&  &0.0836 &  0.1710 & 0.1383    \\ \\ \hline \\

 BLSTM without& 1& 0.0599 & 0.0589   &  0.0529  &0.0503    \\  
 TL&2& 0.0726& 0.0596 &  0.0621 & 0.0711 \\
 ~~~~~&3& 0.0903 &0.0761&  0.0874 & 0.0782   \\ 
 ~~~~~&20& 0.1170 &0.1006 &  0.1205 &0.1183   \\ \\

BLSTM with& 1&  &0.0522 &  0.0476 &0.0548    \\  
 TL&2& &0.0413  & 0.0492 & 0.0683 \\
 ~~~~~&3&  &0.0451&  0.0567 & 0.0696   \\ 
 ~~~~~&20&  &0.0729 & 0.0593 &0.1037    \\ \\ \hline\hline
  \end{tabular}
  \caption{RMSE of LSTM/BLSTM NN without and with TL.}
  \label{tab:tl1}
  \end{center}
\end{table*}

The results show that the effect of TL on the accuracy of the prediction is remarkable as the RMSE has lower values for all the elevations along the obstacle height when the TL is used, compared to the cases when the models are trained without TL, i.e., when the weights are initialized randomly in the training process. It is also shown that when the BLSTM NN is used, the RMSE values are lower than when LSTM NN is used. This can be observed for all the elevations and both of the cases, i.e., without and with TL.\par 

\begin{figure*}
\centering 
\includegraphics[angle=0, trim=0 0 0 0, width=0.8\textwidth]{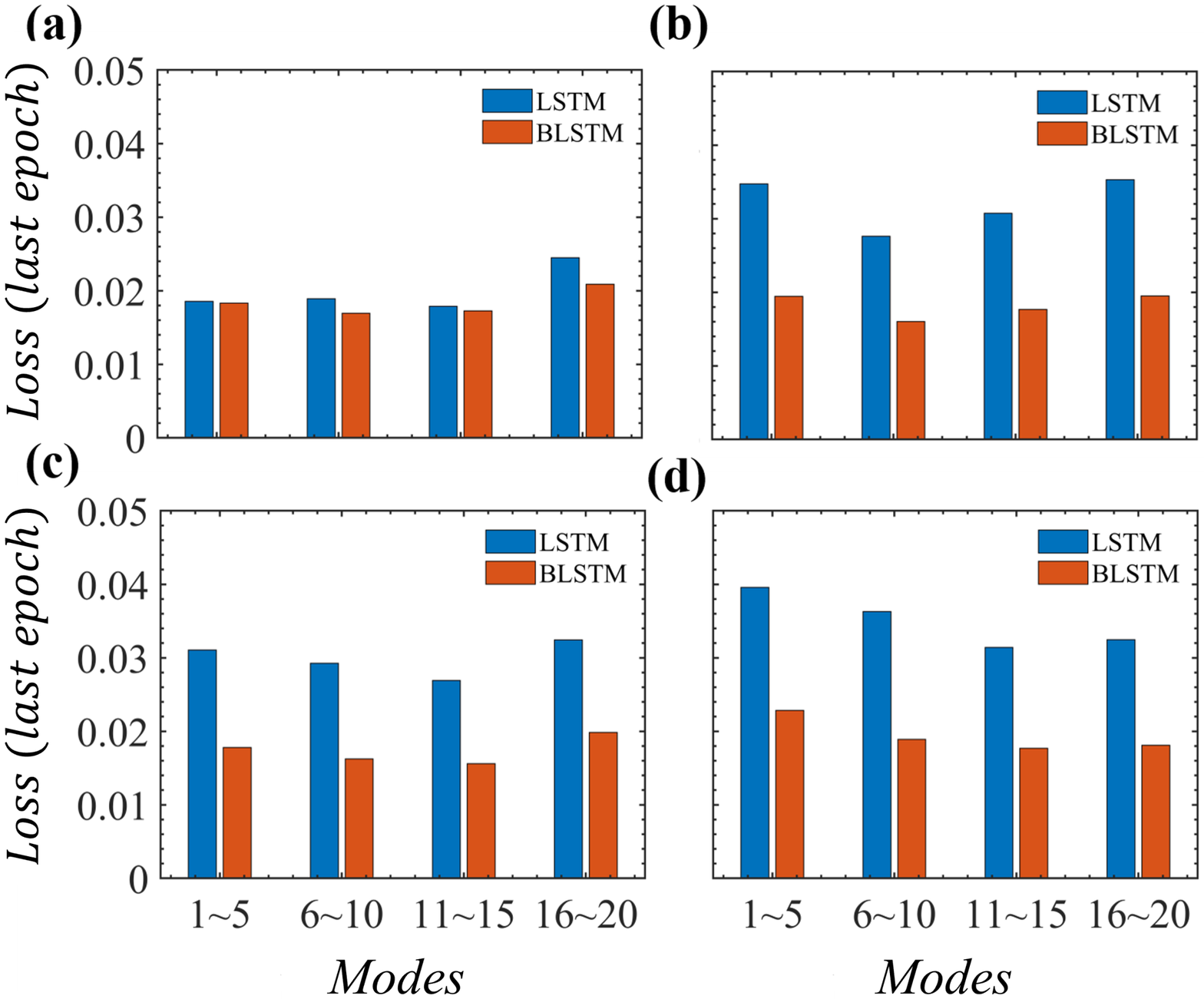}
\caption[]{Comparison of the loss function value of the last epoch at different elevations: (a) $z/D = 0.02$, (b) $z/D = 1$, (c) $z/D = 2$, and  (d) $z/D = 3$.} 
\label{fig:13_loss_bars}
\end{figure*}

Figure~\ref{fig:13_loss_bars} shows a comparison of the loss function value of the last epoch in the training process of LSTM and BLSTM NNs at different elevations along the obstacle. Here the loss function is measured using mean-square error (MSE). It can be seen that the BLSTM NN outperformed the LSTM NN as it keeps the loss function with approximately fixed values even for higher elevations. However, for LSTM NN, the loss function values are affected by the increasing of elevation as the time coefficients exhibit more complex behavior because of the three-dimensional complex nature associated with the shedding process.

\subsection{Prediction of time coefficients and flow field reconstruction}

\begin{figure*}
\centering 
\includegraphics[angle=0, trim=0 0 0 0, width=0.68\textwidth]{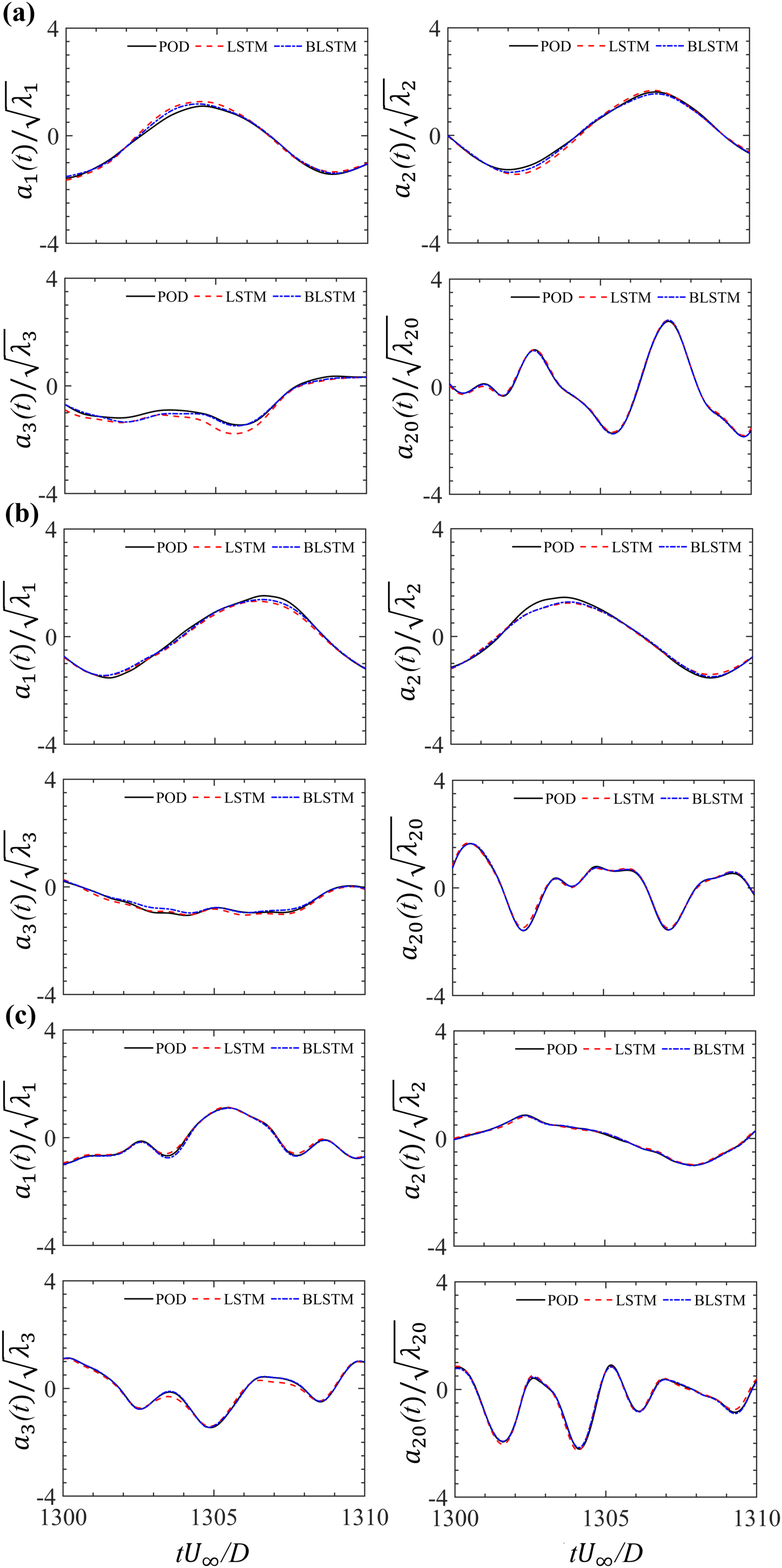}
\caption[]{Plots of the real and predicted time coefficients for different POD modes at different elevations: (a) $z/D = 0.02$, (b) $z/D = 1$, and (c) $z/D = 3$.} 
\label{fig:14_time_coef_plot}
\end{figure*}

One of the essential requirements in this study is to obtain an accurate prediction of the POD time coefficients that can be used along with the corresponding POD modes to reconstruct the flow field. Figure~\ref{fig:14_time_coef_plot} shows a plot of the real and predicted time coefficients for different POD modes along the obstacle height. The predicted time coefficients using LSTM and BLSTM NNs show accuracy similar to the real-time coefficients. Nevertheless, BLSTM NN showes a better prediction for the modes that exhibit complex non-linear behavior. This demonstrates that BLSTM NN has a great potential for addressing complex non-linear time-series problems. After the time coefficients are predicted, the instantaneous reduced-order flow field can be reconstructed using Eq.(\ref{EQ3}). \par

\begin{figure*}
\centering 
\includegraphics[angle=0, trim=0 0 0 0, width=0.7\textwidth]{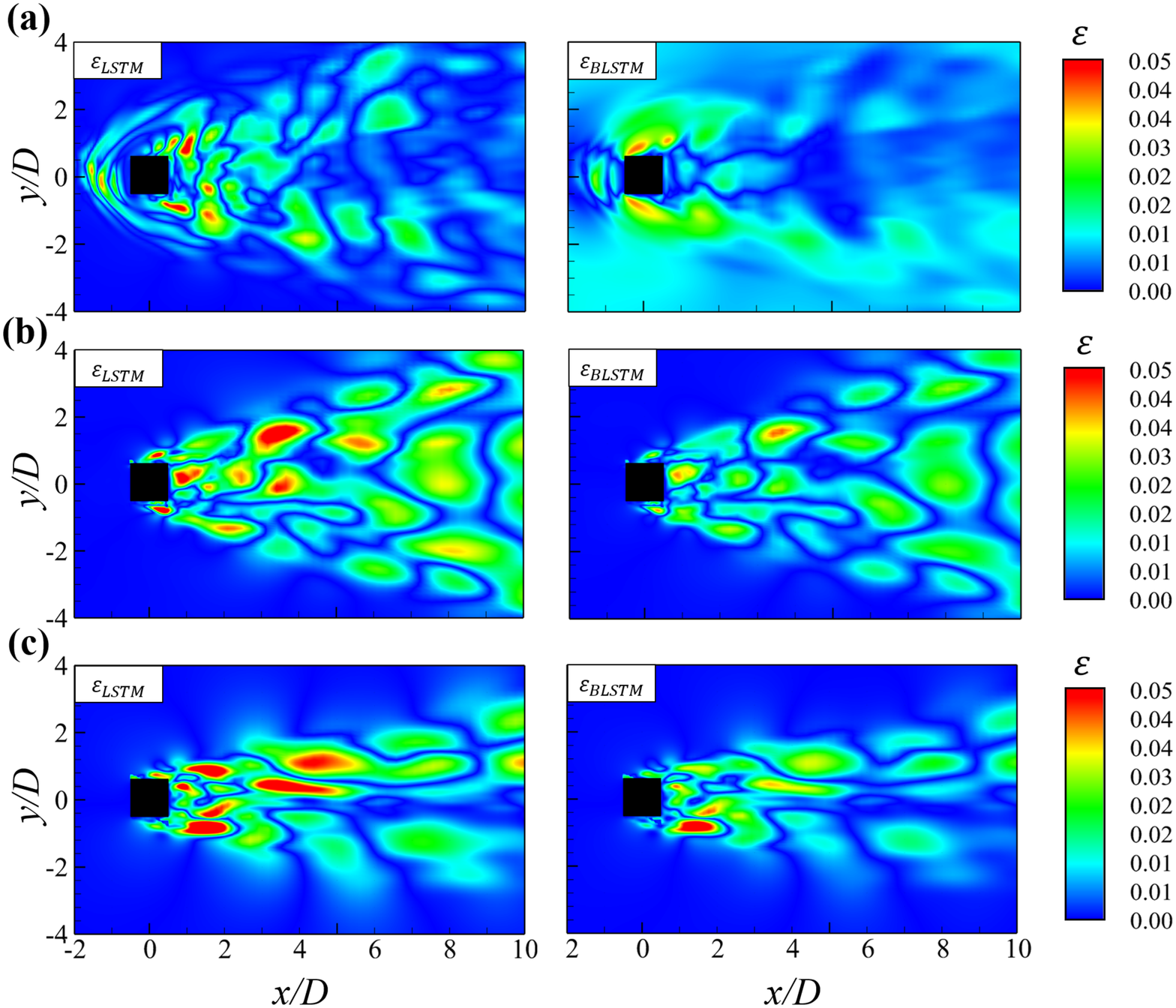}
\caption[]{Contour plots of the absolute reconstruction error of LSTM-POD and BLSTM-POD models at different elevations: (a) $z/D = 0.02$, (b) $z/D = 1$, and (c) $z/D = 3$.} 
\label{fig:16_reconstruction_error}
\end{figure*}

Figure~\ref{fig:15_velocity_reconstruction} shows the contour plots of a random instantaneous streamwise velocity field from the test data calculated by the IDDES and the corresponding reconstructed reduced-order stream-wise velocity field calculated using the POD with real-time coefficients, LSTM-POD model, and BLSTM-POD model, respectively. It is observed from the figure that the first 20 POD modes with the real-time coefficients could successfully reconstruct the large-scale flow structure. However, a slight difference could be seen between the real and the POD reduced-order flow fields in terms of the reconstruction of small-scale eddies. This difference occurs due to the energy-loss because the POD reconstruction considers only 20 modes corresponding to 82\%–92\% of the total turbulent kinetic energy. The velocity field that is reconstructed using BLSTM-POD models demonstrates noticeably better results thanthe one obtained from the LSTM-POD model. As can be obseved from the figure , a slight over-prediction can be recognized from the LSTM-POD contour plots with the increasing elevation along the obstacle. To quantify the reconstruction error of the LSTM-POD and BLSTM-POD models, the absolute reconstruction error was applied as follow:

\begin{figure*}
\centering 
\includegraphics[angle=0, trim=0 0 0 0, width=0.7\textwidth]{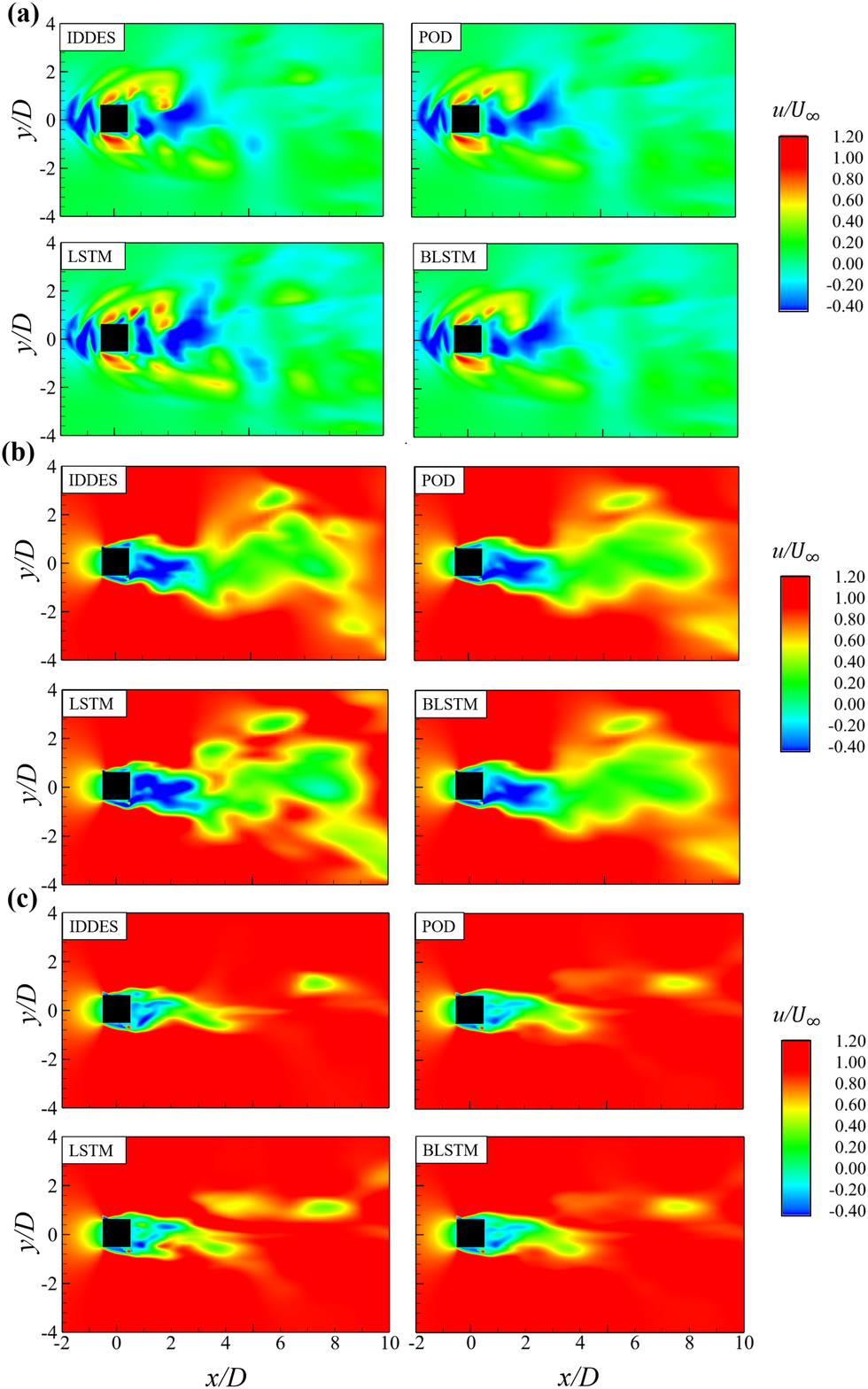}
\caption[]{Contour plots of a random instantaneous streamwise velocity at different elevations: (a) $z/D = 0.02$, (b) $z/D = 1$, and (c) $z/D = 3$.} 
\label{fig:15_velocity_reconstruction}
\end{figure*}

\begin{equation}\label{EQ15}
\varepsilon = \frac{\left | u_{pred} - u_{POD} \right |}{u_\infty},
\end{equation}

\noindent where $u_{POD}$ and $u_{pred}$ are the velocity reconstructed using real POD time coefficients and the time coefficients predicted using LSTM-POD and BLSTM-POD models. As shown in Fig.~\ref{fig:16_reconstruction_error}, the maximum reconstruction error occurs in the vicinity of the large-scale structures in the flow fields, i.e., near the edge of the obstacle and where the shedding happens directly behind the obstacle. The error obtained from the BLSTM-POD model is generally lower than that from the LSM-POD model, which indicates that the BLSTM model outperformed the LSTM model. These results are consistent with the results obtained from Table II and Fig.~\ref{fig:13_loss_bars}.
  
\begin{figure*}
\centering 
\includegraphics[angle=0, trim=0 0 0 0, width=0.8\textwidth]{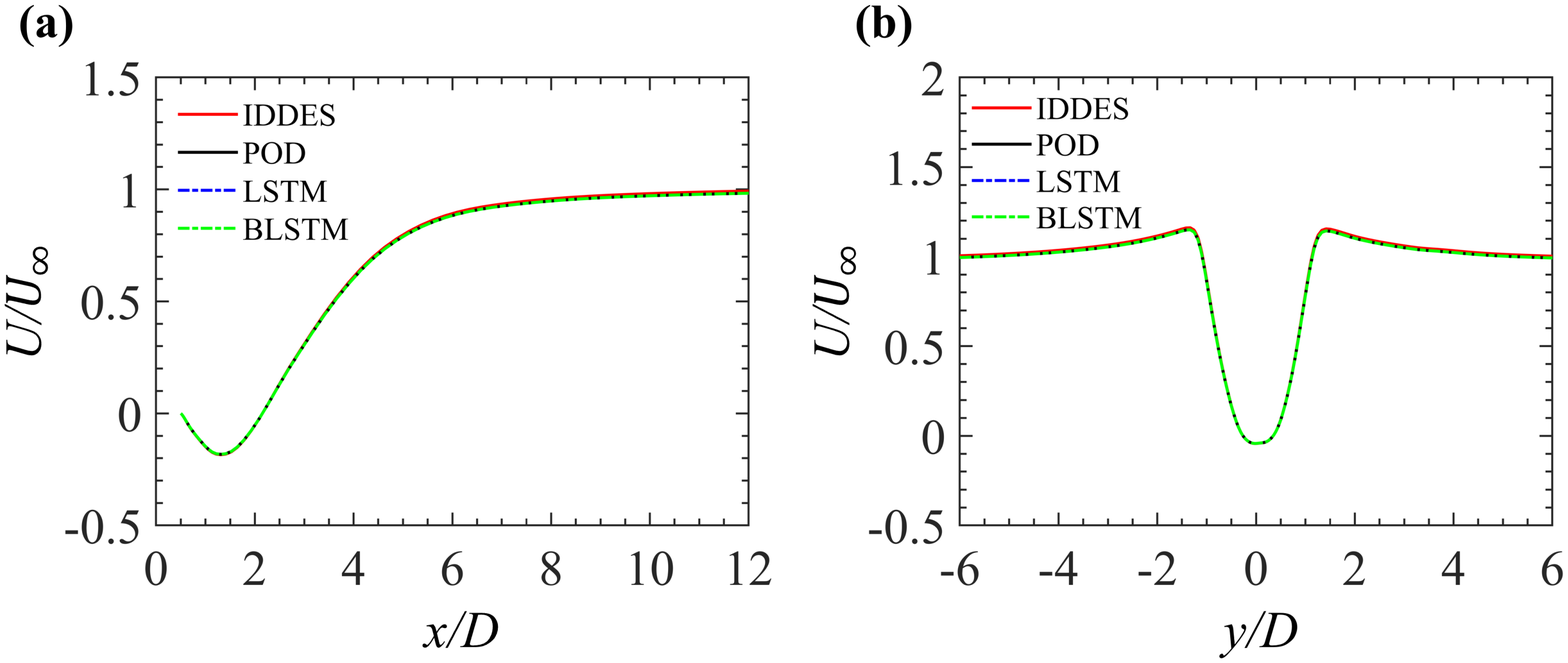}
\caption[]{Streamwise profile of the mean streamwise velocity at elevation $z/D = 3$ in the central $x-z$ plane ($y/D = 0$) (a) and spanwise profile of the mean streamwise velocity at elevation $z/D = 3$ and streamwise location $x = 2$ (b).} 
\label{fig:17_velocity_dl}
\end{figure*}

The two models are used to predict the instantaneous velocity field for 50 000 timesteps that are enough to reproduce the mean velocity profiles as shown in Fig.~\ref{fig:17_velocity_dl} . The results from LSTM and BLSTM-POD models show an excellent agreement with the POD and IDDES results indicating that the LSTM and BLSTM-POD models could successfully capture the main flow fields features and for a sufficiently large number of time steps that can produce a statistically stationary data.

\subsection{Effect of the prediction time window}

LSTM NN is designed to model sequential data using correlations between subsequent realizations. Hence, LSTM NN assume that the realizations of the data are originally correlated with each other. In other words, they share specific features that can together represent the sequential system using the network parameters. This assumption could also be considered as a memory that the network can obtain from the behavior of the sequential data. This memory relies on the characteristics of the sequential data, which can have a periodic or repetitive behavior that can be considered as a long-time correlated event or as chaotic dynamic behavior. Hence, it is important to study the temporal behavior of the POD time coefficients before investigating the effect of the prediction time window length. \par

\begin{figure*}
\centering 
\includegraphics[angle=0, trim=0 0 0 0, width=1.0\textwidth]{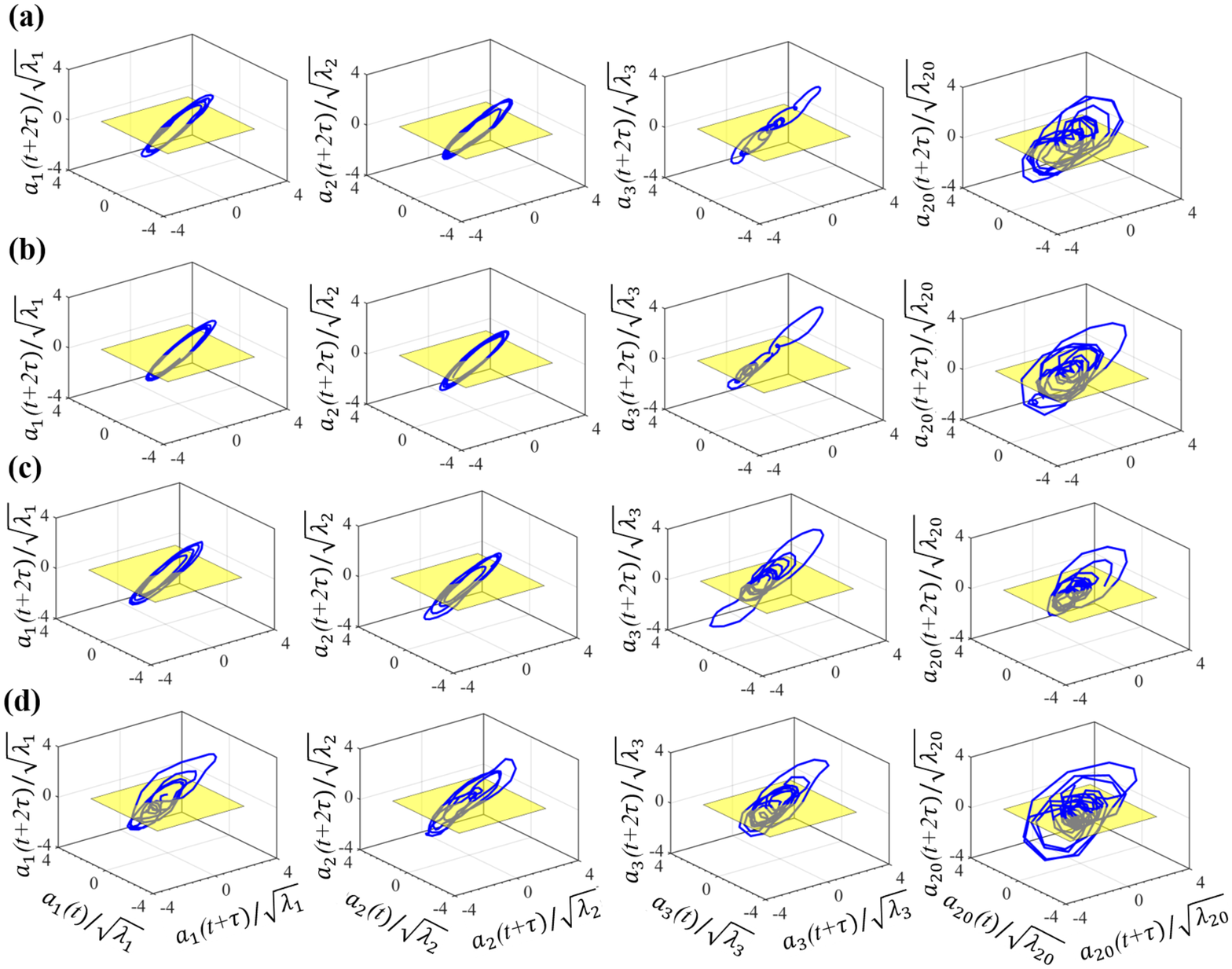}
\caption[]{Phase space plots and Poincar$\acute{e}$ sections of different normalized POD time coefficients at different elevations: (a) z/D = 0.02, (b) z/D = 1, (c) z/D = 2, (d) z/D = 3..} 
\label{fig:18_phase_plot}
\end{figure*}

Fig.~\ref{fig:18_phase_plot} shows the phase space plots of the time coefficients, where $\tau$ is the time lag, which is considered as 10 time-steps. The yellow plane in each plot illustrates the Poincar$\acute{e}$ section \cite{Cavers2017}. Poincar$\acute{e}$ section shows the intersection points of the trajectories of the discrete time interval with an infinite plane. The set of the intersection points on this plane indicates if the dynamical system is periodic, quasi-periodic, or has a chaotic behavior. For instance, for a pure periodic system, two distinct points will appear on the section, while for a quasi-periodic system, the points will be scattered in a certain deterministic order. For systems with chaotic behavior, the points will be scattered with irregular patterns. As can be seen from the figure, at elevations $z/D$ = 0.02, 1 and 2, the phase space plots of the first two time coefficients show a quasi-periodic behavior with a slight decrease in the periodicity as the elevation increases. However, at $z/D$ = 3, this behavior is barely recognized due to the aforementioned downwash effect of the obstacle-free end. \par

For the remaining time coefficients, chaotic behavior can be recognized, indicating that these coefficients exhibit lesser long-term correlations compared to the first two time coefficients. Considering these observations, LSTM and BLSTM NNs are trained using two different sample lengths of 50 and 100 time steps to predict the time coefficients with time windows of 50 and 100 time steps, respectively. \par

Table III shows the RMSE of the predicted time coefficients using time window lengths of 50 and 100 time steps with the percentage increase of the RMSE compared to a predicted time window of 10 time steps. In general, the RMSE shows higher values as the prediction time window lengths increase. This reduction in the accuracy can be explained by considering {\it Lyapunov exponent theory} \cite{Wolfetal1985} , which states that as the trajectories of a dynamic system move farther in time from the initial state, a continuous divergence starts emerging between these trajectories. As expected, the time coefficients with a higher POD ranking show higher RMSE values compared to the time coefficients that correspond to the first two POD modes. This indicates that the long-term memory is reduced as the POD ranking of the time coefficients increases, which is consistent with the results obtained from Fig.~\ref{fig:18_phase_plot}. Interestingly, the percentage rate of increase of RMSE reduces as the POD ranking increases indicating that even though they have a long-term correlation, the time coefficients with low POD ranking tend to demonstrate a faster reduction in the prediction accuracy as the prediction time window increases.\par

\begin{table*}
  \begin{center}
  \begin{tabular}{c c  c  c  c c c}\hline\hline \\
  Prediction time window (time steps)&~~~~~~& POD mode & $z/D = 0.02 $ &  $z/d = 1$ & $z/D = 2$ &$ z/D = 3$ \\[3pt] \hline \\
50& LSTM& 1& 0.1955 (211.3\%)&0.1714 (186.1\%)   &  0.1865 (146.0\%) & 0.2623 (293.8\%) \\  
 &&2&0.2627 (177.1\%) & 0.1761 (154.1\%) &  0.2663 (190.7\%) & 0.2805 (289.0\%) \\
& ~~~~~&3& 0.2560 (165.8\%)&0.1943 (135.5\%) &  0.3174 (173.1\%)& 0.3059 (281.4\%)   \\ 
 &~~~~~&20& 0.2473 (92.4\%)&0.2198 (99.8\%)& 0.4283 (147.4\%) & 0.4271 (191.1\%)   \\ \\

&BLSTM &1& 0.1543(157.5\%)&0.1692 (187.2\%)& 0.1581 (198.8\%)& 0.2014 (300.3\%)\\ 
&&2& 0.1712 (135.8\%)  &  0.1693 (184.0\%)  & 0.1851 (198.0\%)&0.2544 (257.8\%)\\
& ~~~~~&3& 0.2115 (134.2\%)&  0.1906 (131.0\%) & 0.2512 (187.4\%)&0.2804 (258.5\%)  \\ 
 &~~~~~&20&0.2352 (83.0\%)&0.2088 (107.5\%)& 0.3451 (186.3\%)&0.3808 (221.8\%)  \\ \\ \hline \\

100&LSTM & 1& 0.2153 (208.2\%) & 0.2251 (275.8\%)  &  0.3898 (414.2\%) &0.4629 (595.0\%)    \\  
 &&2&0.2922 (203.4\%)& 0.2276 (228.4\%) & 0.4651 (407.7\%)& 0.4972 (589.6\%) \\
& ~~~~~&3& 0.3101 (222.0\%) &0.2322 (181.4\%)&0.4932 (324.4\%) & 0.5516 (587.7\%)  \\ 
 &~~~~~&20& 0.3928 (205.7\%) &0.2819 (156.2\%) &0.5243 (202.8\%) &0.6316 (330.5\%)   \\ \\

&BLSTM & 1&  0.1974 (229.5) &0.2168 (268.0\%) &0.3214 (507.5\%)&0.4015 (698.2\%)    \\  
&&2& 0.2361 (225.2\%) & 0.1645 (176.0\%)& 0.3631 (484.7\%)&0.4692 (559.9\%) \\
 &~~~~~&3& 0.2675 (1962.2\%)&0.2032 (146.0\%) & 0.4758 (444.3\%)&0.5074 (548.8\%)  \\ 
 &~~~~~&20&0.3135 (143.9\%)& 0.2512 (149.7\%)&0.5001 (315.0\%)&0.6173 (421.8\%)    \\ \\ \hline\hline
  \end{tabular}
  \caption{ RMSE of LSTM/BLSTM NN and percentage rate (in parentheses) of RMSE increasing with the use of two different prediction time windows.}
  \label{tab:tl2}
  \end{center}
\end{table*}

\section{Conclusion}
This study presented a ROM based on LSTM/BLSTM NN and POD that can predict the instantaneous turbulent wake behavior of a finite wall-mounted square cylinder. The main idea of the ROM in this study is to accurately predict the POD time coefficients for several future time steps instantaneously using the LSTM/BLSTM NN and reconstruct the flow field using a few selected POD modes. The following summarizes the conclusions and findings of this study:

\begin{enumerate}
\item Because the first 20 POD modes for all the x-y planes along the obstacle height contribute to 82\%–92\% of the total turbulent kinetic energy, they can successfully reconstruct the main features of the flow field with the corresponding time coefficients. Nevertheless, an inevitable difference is recognized between the reduced-order flow fields and the flow obtained from the IDDES due to the energy loss of the POD reconstruction. 

\item The use of the TL approach in this study shows a noticeable improvement in the accuracy of the prediction. 

\item A comparative study of LSTM and BLSTM NNs revealed that BLSTM NN outperforms LSTM NN in both training and prediction accuracy, suggesting that BLSTM NN has more potential to predict time series data with more complex nonlinear behavior comparing with LSTM NN.

\item Both LSTM-POD and BLSTM-POD ROMs demonstrate good reconstruction results for most of the regions in the flow fields. However, the reconstruction error analysis showed that the BLSTM-POD ROM outperforms the LSTM-POD ROM at the regions near the obstacle edges and right behind the obstacle, which is consistent with the results obtained from the training and prediction error analysis. 

\item The flow predicted by LSTM-POD and BLSTM-POD ROMs showed a statistically good agreement with the reference IDDES data, suggesting that the ROMs could predict the main features of the flow and for a long time that is sufficient to produce statistically stationary data.

\item The phase plots and Poincar$\acute{e}$ sections of the POD time coefficients showed that the periodicity of the flow is mainly represented by the first two POD modes and it decreases with increasing obstacle elevation, indicating that the downwash effect from the obstacle-free end becomes more dominant. For the rest of the POD time coefficients, a chaotic behavior can be seen. 

\item The results of training the LSTM/BLSTM NN with longer lengths of the prediction time window revealed that the time coefficients corresponding to the first two POD modes have more potential for prediction even with longer lengths of the prediction time window because they have long time-correlated behavior. However, as they showed a higher increasing rate of the RMSE as the prediction time window length increases, the time coefficients with lower POD ranking tend to have a faster decrease in the prediction accuracy comparing with the time coefficients of higher POD ranking.

\end{enumerate}

The previous observations indicate that the ROM presented in this study could predict the reduced-order flow field of the turbulent wake with commendable accuracy when the BLSTM-POD model is used. As the prediction accuracy of the model is affected by the increase in the prediction time window length, further studies focusing on the optimization of the model will be conducted to improve the prediction accuracy for longer prediction time windows. Moreover, the prediction capability of the proposed ROM will be further examined using a range of $Re$ and different obstacles shapes.

\begin{acknowledgments}
This work was supported by 'Human Resources Program in Energy Technology' of the Korea Institute of Energy Technology Evaluation and Planning (KETEP), granted financial resource from the Ministry of Trade, Industry \& Energy, Republic of Korea (no. 20184030202200). In addition, this work was supported by the National Research Foundation of Korea (NRF) grant funded by the Korea government (MSIP) (no. 2019R1I1A3A01058576). This work was also supported by the National Supercomputing Center with supercomputing resources including technical support (KSC-2020-INO-0025).
\end{acknowledgments}

\section*{Data Availability}
The data that supports the findings of this study are available within this article.

\appendix

\nocite{*}
\bibliography{my-bib}

\providecommand{\noopsort}[1]{}\providecommand{\singleletter}[1]{#1}%
\begin{thebibliography}{57}%
\makeatletter
\providecommand \@ifxundefined [1]{%
 \@ifx{#1\undefined}
}%
\providecommand \@ifnum [1]{%
 \ifnum #1\expandafter \@firstoftwo
 \else \expandafter \@secondoftwo
 \fi
}%
\providecommand \@ifx [1]{%
 \ifx #1\expandafter \@firstoftwo
 \else \expandafter \@secondoftwo
 \fi
}%
\providecommand \natexlab [1]{#1}%
\providecommand \enquote  [1]{``#1''}%
\providecommand \bibnamefont  [1]{#1}%
\providecommand \bibfnamefont [1]{#1}%
\providecommand \citenamefont [1]{#1}%
\providecommand \href@noop [0]{\@secondoftwo}%
\providecommand \href [0]{\begingroup \@sanitize@url \@href}%
\providecommand \@href[1]{\@@startlink{#1}\@@href}%
\providecommand \@@href[1]{\endgroup#1\@@endlink}%
\providecommand \@sanitize@url [0]{\catcode `\\12\catcode `\$12\catcode
  `\&12\catcode `\#12\catcode `\^12\catcode `\_12\catcode `\%12\relax}%
\providecommand \@@startlink[1]{}%
\providecommand \@@endlink[0]{}%
\providecommand \url  [0]{\begingroup\@sanitize@url \@url }%
\providecommand \@url [1]{\endgroup\@href {#1}{\urlprefix }}%
\providecommand \urlprefix  [0]{URL }%
\providecommand \Eprint [0]{\href }%
\providecommand \doibase [0]{https://doi.org/}%
\providecommand \selectlanguage [0]{\@gobble}%
\providecommand \bibinfo  [0]{\@secondoftwo}%
\providecommand \bibfield  [0]{\@secondoftwo}%
\providecommand \translation [1]{[#1]}%
\providecommand \BibitemOpen [0]{}%
\providecommand \bibitemStop [0]{}%
\providecommand \bibitemNoStop [0]{.\EOS\space}%
\providecommand \EOS [0]{\spacefactor3000\relax}%
\providecommand \BibitemShut  [1]{\csname bibitem#1\endcsname}%
\let\auto@bib@innerbib\@empty
\bibitem [{\citenamefont {Bourgeois}, \citenamefont {Sattari},\ and\
  \citenamefont {Martinuzzi}(2011)}]{Bourgeoisetal2011}%
  \BibitemOpen
  \bibfield  {author} {\bibinfo {author} {\bibfnamefont {J.~A.}\ \bibnamefont
  {Bourgeois}}, \bibinfo {author} {\bibfnamefont {P.}~\bibnamefont {Sattari}},\
  and\ \bibinfo {author} {\bibfnamefont {R.~J.}\ \bibnamefont {Martinuzzi}},\
  }\bibfield  {title} {\enquote {\bibinfo {title} {Alternating half-loop
  shedding in the turbulent wake of a finite surface-mounted square cylinder
  with a thin boundary layer},}\ }\href@noop {} {\bibfield  {journal} {\bibinfo
   {journal} {Physics of Fluids}\ }\textbf {\bibinfo {volume} {23(9)}},\
  \bibinfo {pages} {095101} (\bibinfo {year} {2011})}\BibitemShut {NoStop}%
\bibitem [{\citenamefont {Park}\ and\ \citenamefont
  {Lee}(2000)}]{Park&Lee2000}%
  \BibitemOpen
  \bibfield  {author} {\bibinfo {author} {\bibfnamefont {C.-W.}\ \bibnamefont
  {Park}}\ and\ \bibinfo {author} {\bibfnamefont {S.-J.}\ \bibnamefont {Lee}},\
  }\bibfield  {title} {\enquote {\bibinfo {title} {Free end effects on the near
  wake flow structure behind a finite circular cylinder},}\ }\href@noop {}
  {\bibfield  {journal} {\bibinfo  {journal} {Journal of Wind Engineering and
  Industrial Aerodynamics}\ }\textbf {\bibinfo {volume} {88(2–3)}},\ \bibinfo
  {pages} {231--246} (\bibinfo {year} {2000})}\BibitemShut {NoStop}%
\bibitem [{\citenamefont {Sattari}, \citenamefont {Bourgeois},\ and\
  \citenamefont {Martinuzzi}(2012)}]{Sattarietal2012}%
  \BibitemOpen
  \bibfield  {author} {\bibinfo {author} {\bibfnamefont {P.}~\bibnamefont
  {Sattari}}, \bibinfo {author} {\bibfnamefont {J.~A.}\ \bibnamefont
  {Bourgeois}},\ and\ \bibinfo {author} {\bibfnamefont {R.~J.}\ \bibnamefont
  {Martinuzzi}},\ }\bibfield  {title} {\enquote {\bibinfo {title} {On the
  vortex dynamics in the wake of a finite surface-mounted square cylinder},}\
  }\href@noop {} {\bibfield  {journal} {\bibinfo  {journal} {Experiments in
  Fluids}\ }\textbf {\bibinfo {volume} {52(5)}},\ \bibinfo {pages} {1149--1167}
  (\bibinfo {year} {2012})}\BibitemShut {NoStop}%
\bibitem [{\citenamefont {Saeedi}, \citenamefont {LePoudre},\ and\
  \citenamefont {Wang}(2014)}]{Saeedietal2014}%
  \BibitemOpen
  \bibfield  {author} {\bibinfo {author} {\bibfnamefont {M.}~\bibnamefont
  {Saeedi}}, \bibinfo {author} {\bibfnamefont {P.~P.}\ \bibnamefont
  {LePoudre}},\ and\ \bibinfo {author} {\bibfnamefont {B.~C.}\ \bibnamefont
  {Wang}},\ }\bibfield  {title} {\enquote {\bibinfo {title} {Direct numerical
  simulation of turbulent wake behind a surface-mounted square cylinder},}\
  }\href@noop {} {\bibfield  {journal} {\bibinfo  {journal} {Journal of Fluids
  and Structures}\ }\textbf {\bibinfo {volume} {50}},\ \bibinfo {pages}
  {20--39} (\bibinfo {year} {2014})}\BibitemShut {NoStop}%
\bibitem [{\citenamefont {Tang}, \citenamefont {Graham},\ and\ \citenamefont
  {Peraire}(1996)}]{Tangetal1996}%
  \BibitemOpen
  \bibfield  {author} {\bibinfo {author} {\bibfnamefont {K.}~\bibnamefont
  {Tang}}, \bibinfo {author} {\bibfnamefont {W.}~\bibnamefont {Graham}},\ and\
  \bibinfo {author} {\bibfnamefont {J.}~\bibnamefont {Peraire}},\ }\bibfield
  {title} {\enquote {\bibinfo {title} {Active flow control using a reduced
  order model and optimum control},}\ }\href@noop {} {\bibfield  {journal}
  {\bibinfo  {journal} {https://doi.org/10.2514/6.1996-1946}\ } (\bibinfo
  {year} {1996})}\BibitemShut {NoStop}%
\bibitem [{\citenamefont {Wang}\ and\ \citenamefont
  {Zhou}(2009)}]{Wang&Zhou2009}%
  \BibitemOpen
  \bibfield  {author} {\bibinfo {author} {\bibfnamefont {H.~F.}\ \bibnamefont
  {Wang}}\ and\ \bibinfo {author} {\bibfnamefont {Y.}~\bibnamefont {Zhou}},\
  }\bibfield  {title} {\enquote {\bibinfo {title} {The finite-length square
  cylinder near wake},}\ }\href@noop {} {\bibfield  {journal} {\bibinfo
  {journal} {Journal of Fluid Mechanics}\ }\textbf {\bibinfo {volume} {638}},\
  \bibinfo {pages} {453--490} (\bibinfo {year} {2009})}\BibitemShut {NoStop}%
\bibitem [{\citenamefont {Yousif}\ and\ \citenamefont
  {Lim}(2020)}]{Yousif&Lim2020}%
  \BibitemOpen
  \bibfield  {author} {\bibinfo {author} {\bibfnamefont {M.~Z.~G.}\
  \bibnamefont {Yousif}}\ and\ \bibinfo {author} {\bibfnamefont
  {H.}~\bibnamefont {Lim}},\ }\bibfield  {title} {\enquote {\bibinfo {title}
  {On the characteristics of the turbulent wake behind a wall-mounted square
  cylinder},}\ }\href@noop {} {\bibfield  {journal} {\bibinfo  {journal}
  {ArXiv:2012.11263 [Physics].}\ } (\bibinfo {year} {2020})}\BibitemShut
  {NoStop}%
\bibitem [{\citenamefont {Lumley}(1967)}]{Lumley1967}%
  \BibitemOpen
  \bibfield  {author} {\bibinfo {author} {\bibfnamefont {J.~L.}\ \bibnamefont
  {Lumley}},\ }\bibfield  {title} {\enquote {\bibinfo {title} {The structure of
  inhomogeneous turbulent flows},}\ }\href@noop {} {\bibfield  {journal}
  {\bibinfo  {journal} {Atmospheric Turbulence and Radio Wave Propagation}\ }
  (\bibinfo {year} {1967})}\BibitemShut {NoStop}%
\bibitem [{\citenamefont {Schmid}(2010)}]{Schmid2010}%
  \BibitemOpen
  \bibfield  {author} {\bibinfo {author} {\bibfnamefont {P.~J.}\ \bibnamefont
  {Schmid}},\ }\bibfield  {title} {\enquote {\bibinfo {title} {Dynamic mode
  decomposition of numerical and experimental data},}\ }\href@noop {}
  {\bibfield  {journal} {\bibinfo  {journal} {Journal of Fluid Mechanics}\
  }\textbf {\bibinfo {volume} {656}},\ \bibinfo {pages} {5--28} (\bibinfo
  {year} {2010})}\BibitemShut {NoStop}%
\bibitem [{\citenamefont {Burkardt}, \citenamefont {Gunzburger},\ and\
  \citenamefont {Lee}(2006)}]{Burkardtetal2006}%
  \BibitemOpen
  \bibfield  {author} {\bibinfo {author} {\bibfnamefont {J.}~\bibnamefont
  {Burkardt}}, \bibinfo {author} {\bibfnamefont {M.}~\bibnamefont
  {Gunzburger}},\ and\ \bibinfo {author} {\bibfnamefont {H.-C.}\ \bibnamefont
  {Lee}},\ }\bibfield  {title} {\enquote {\bibinfo {title} {Pod and cvt-based
  reduced-order modeling of navier–stokes flows},}\ }\href@noop {} {\bibfield
   {journal} {\bibinfo  {journal} {Computer Methods in Applied Mechanics and
  Engineering}\ }\textbf {\bibinfo {volume} {196(1)}},\ \bibinfo {pages}
  {337--355} (\bibinfo {year} {2006})}\BibitemShut {NoStop}%
\bibitem [{\citenamefont {Carlberg}, \citenamefont {Bou-Mosleh},\ and\
  \citenamefont {Farhat}(2011)}]{Carlbergetal2011}%
  \BibitemOpen
  \bibfield  {author} {\bibinfo {author} {\bibfnamefont {K.}~\bibnamefont
  {Carlberg}}, \bibinfo {author} {\bibfnamefont {C.}~\bibnamefont
  {Bou-Mosleh}},\ and\ \bibinfo {author} {\bibfnamefont {C.}~\bibnamefont
  {Farhat}},\ }\bibfield  {title} {\enquote {\bibinfo {title} {Efficient
  non-linear model reduction via a least-squares petrov–galerkin projection
  and compressive tensor approximations},}\ }\href@noop {} {\bibfield
  {journal} {\bibinfo  {journal} {International Journal for Numerical Methods
  in Engineering}\ }\textbf {\bibinfo {volume} {86(2)}},\ \bibinfo {pages}
  {155--181} (\bibinfo {year} {2011})}\BibitemShut {NoStop}%
\bibitem [{\citenamefont {Rap$\acute{u}$n}\ and\ \citenamefont
  {Vega}(2010)}]{Rapun&Vega2010}%
  \BibitemOpen
  \bibfield  {author} {\bibinfo {author} {\bibfnamefont {M.-L.}\ \bibnamefont
  {Rap$\acute{u}$n}}\ and\ \bibinfo {author} {\bibfnamefont {J.~M.}\
  \bibnamefont {Vega}},\ }\bibfield  {title} {\enquote {\bibinfo {title}
  {Reduced order models based on local pod plus galerkin projection},}\
  }\href@noop {} {\bibfield  {journal} {\bibinfo  {journal} {Journal of
  Computational Physics}\ }\textbf {\bibinfo {volume} {229(8)}},\ \bibinfo
  {pages} {3046--3063} (\bibinfo {year} {2010})}\BibitemShut {NoStop}%
\bibitem [{\citenamefont {Rowley}, \citenamefont {Colonius},\ and\
  \citenamefont {Murray}(2000)}]{Rowleyetal2000}%
  \BibitemOpen
  \bibfield  {author} {\bibinfo {author} {\bibfnamefont {C.~W.}\ \bibnamefont
  {Rowley}}, \bibinfo {author} {\bibfnamefont {T.}~\bibnamefont {Colonius}},\
  and\ \bibinfo {author} {\bibfnamefont {R.~M.}\ \bibnamefont {Murray}},\
  }\bibfield  {title} {\enquote {\bibinfo {title} {Model reduction for
  compressible flows using pod and galerkin projection},}\ }\href@noop {}
  {\bibfield  {journal} {\bibinfo  {journal} {Physica D: Nonlinear Phenomena}\
  }\textbf {\bibinfo {volume} {189(1)}},\ \bibinfo {pages} {115--129} (\bibinfo
  {year} {2000})}\BibitemShut {NoStop}%
\bibitem [{\citenamefont {Akhtar}, \citenamefont {Nayfeh},\ and\ \citenamefont
  {Ribbens}(2009)}]{Akhtaretal2009}%
  \BibitemOpen
  \bibfield  {author} {\bibinfo {author} {\bibfnamefont {I.}~\bibnamefont
  {Akhtar}}, \bibinfo {author} {\bibfnamefont {A.~H.}\ \bibnamefont {Nayfeh}},\
  and\ \bibinfo {author} {\bibfnamefont {C.~J.}\ \bibnamefont {Ribbens}},\
  }\bibfield  {title} {\enquote {\bibinfo {title} {On the stability and
  extension of reduced-order galerkin models in incompressible flows.}}\
  }\href@noop {} {\bibfield  {journal} {\bibinfo  {journal} {Theoretical and
  Computational Fluid Dynamics}\ }\textbf {\bibinfo {volume} {23(3)}},\
  \bibinfo {pages} {213--237} (\bibinfo {year} {2009})}\BibitemShut {NoStop}%
\bibitem [{\citenamefont {Rempfer}(2000)}]{Rempfer2000}%
  \BibitemOpen
  \bibfield  {author} {\bibinfo {author} {\bibfnamefont {D.}~\bibnamefont
  {Rempfer}},\ }\bibfield  {title} {\enquote {\bibinfo {title} {On
  low-dimensional galerkin models for fluid flow},}\ }\href@noop {} {\bibfield
  {journal} {\bibinfo  {journal} {Theoretical and Computational Fluid
  Dynamics}\ }\textbf {\bibinfo {volume} {14(2)}},\ \bibinfo {pages} {75--88}
  (\bibinfo {year} {2000})}\BibitemShut {NoStop}%
\bibitem [{\citenamefont {Shinde}\ \emph {et~al.}(2016)\citenamefont {Shinde},
  \citenamefont {Longatte}, \citenamefont {Baj}, \citenamefont {Hoarau},\ and\
  \citenamefont {Braza}}]{Shindeetal2016}%
  \BibitemOpen
  \bibfield  {author} {\bibinfo {author} {\bibfnamefont {V.}~\bibnamefont
  {Shinde}}, \bibinfo {author} {\bibfnamefont {E.}~\bibnamefont {Longatte}},
  \bibinfo {author} {\bibfnamefont {F.}~\bibnamefont {Baj}}, \bibinfo {author}
  {\bibfnamefont {Y.}~\bibnamefont {Hoarau}},\ and\ \bibinfo {author}
  {\bibfnamefont {M.}~\bibnamefont {Braza}},\ }\bibfield  {title} {\enquote
  {\bibinfo {title} {A galerkin-free model reduction approach for the
  navier–stokes equations},}\ }\href@noop {} {\bibfield  {journal} {\bibinfo
  {journal} {Journal of Computational Physics}\ }\textbf {\bibinfo {volume}
  {309}},\ \bibinfo {pages} {148--163} (\bibinfo {year} {2016})}\BibitemShut
  {NoStop}%
\bibitem [{\citenamefont {LeCun}, \citenamefont {Bengio},\ and\ \citenamefont
  {Hinton}(2015)}]{LeCunetal2015}%
  \BibitemOpen
  \bibfield  {author} {\bibinfo {author} {\bibfnamefont {Y.}~\bibnamefont
  {LeCun}}, \bibinfo {author} {\bibfnamefont {Y.}~\bibnamefont {Bengio}},\ and\
  \bibinfo {author} {\bibfnamefont {G.}~\bibnamefont {Hinton}},\ }\bibfield
  {title} {\enquote {\bibinfo {title} {Deep learning},}\ }\href@noop {}
  {\bibfield  {journal} {\bibinfo  {journal} {Nature}\ }\textbf {\bibinfo
  {volume} {521(7553)}},\ \bibinfo {pages} {436--444} (\bibinfo {year}
  {2015})}\BibitemShut {NoStop}%
\bibitem [{\citenamefont {Brunton}, \citenamefont {Noack},\ and\ \citenamefont
  {Koumoutsakos}(2020)}]{Bruntonetal2020}%
  \BibitemOpen
  \bibfield  {author} {\bibinfo {author} {\bibfnamefont {S.~L.}\ \bibnamefont
  {Brunton}}, \bibinfo {author} {\bibfnamefont {B.~R.}\ \bibnamefont {Noack}},\
  and\ \bibinfo {author} {\bibfnamefont {P.}~\bibnamefont {Koumoutsakos}},\
  }\bibfield  {title} {\enquote {\bibinfo {title} {Machine learning for fluid
  mechanics},}\ }\href@noop {} {\bibfield  {journal} {\bibinfo  {journal}
  {Annual Review of Fluid Mechanics}\ }\textbf {\bibinfo {volume} {52(1)}},\
  \bibinfo {pages} {477--508} (\bibinfo {year} {2020})}\BibitemShut {NoStop}%
\bibitem [{\citenamefont {Kutz}(2017)}]{Kutz2017}%
  \BibitemOpen
  \bibfield  {author} {\bibinfo {author} {\bibfnamefont {J.~N.}\ \bibnamefont
  {Kutz}},\ }\bibfield  {title} {\enquote {\bibinfo {title} {Deep learning in
  fluid dynamics},}\ }\href@noop {} {\bibfield  {journal} {\bibinfo  {journal}
  {Journal of Fluid Mechanics}\ }\textbf {\bibinfo {volume} {814}},\ \bibinfo
  {pages} {1--4} (\bibinfo {year} {2017})}\BibitemShut {NoStop}%
\bibitem [{\citenamefont {Baldi}\ and\ \citenamefont
  {Hornik}(1989)}]{Baldi&Hornik1989}%
  \BibitemOpen
  \bibfield  {author} {\bibinfo {author} {\bibfnamefont {P.}~\bibnamefont
  {Baldi}}\ and\ \bibinfo {author} {\bibfnamefont {K.}~\bibnamefont {Hornik}},\
  }\bibfield  {title} {\enquote {\bibinfo {title} {Neural networks and
  principal component analysis: Learning from examples without local minima},}\
  }\href@noop {} {\bibfield  {journal} {\bibinfo  {journal} {Neural Networks}\
  }\textbf {\bibinfo {volume} {2(1)}},\ \bibinfo {pages} {53--58} (\bibinfo
  {year} {1989})}\BibitemShut {NoStop}%
\bibitem [{\citenamefont {Bourlard}\ and\ \citenamefont
  {Kamp}(1988)}]{Bourlard&Kamp1988}%
  \BibitemOpen
  \bibfield  {author} {\bibinfo {author} {\bibfnamefont {H.}~\bibnamefont
  {Bourlard}}\ and\ \bibinfo {author} {\bibfnamefont {Y.}~\bibnamefont
  {Kamp}},\ }\bibfield  {title} {\enquote {\bibinfo {title} {Auto-association
  by multilayer perceptrons and singular value decomposition},}\ }\href@noop {}
  {\bibfield  {journal} {\bibinfo  {journal} {Biological Cybernetics}\ }\textbf
  {\bibinfo {volume} {54(4)}},\ \bibinfo {pages} {291--294} (\bibinfo {year}
  {1988})}\BibitemShut {NoStop}%
\bibitem [{\citenamefont {Milano}\ and\ \citenamefont
  {Koumoutsakos}(2002)}]{Milano&Koumoutsakos2002}%
  \BibitemOpen
  \bibfield  {author} {\bibinfo {author} {\bibfnamefont {M.}~\bibnamefont
  {Milano}}\ and\ \bibinfo {author} {\bibfnamefont {P.}~\bibnamefont
  {Koumoutsakos}},\ }\bibfield  {title} {\enquote {\bibinfo {title} {Neural
  network modeling for near wall turbulent flow},}\ }\href@noop {} {\bibfield
  {journal} {\bibinfo  {journal} {Journal of Computational Physics}\ }\textbf
  {\bibinfo {volume} {182(1)}},\ \bibinfo {pages} {1--26} (\bibinfo {year}
  {2002})}\BibitemShut {NoStop}%
\bibitem [{\citenamefont {G$\ddot{u}$emes}, \citenamefont {Discetti},\ and\
  \citenamefont {Ianiro}(2019)}]{Guemesetal2019}%
  \BibitemOpen
  \bibfield  {author} {\bibinfo {author} {\bibfnamefont {A.}~\bibnamefont
  {G$\ddot{u}$emes}}, \bibinfo {author} {\bibfnamefont {S.}~\bibnamefont
  {Discetti}},\ and\ \bibinfo {author} {\bibfnamefont {A.}~\bibnamefont
  {Ianiro}},\ }\bibfield  {title} {\enquote {\bibinfo {title} {Sensing the
  turbulent large-scale motions with their wall signature},}\ }\href@noop {}
  {\bibfield  {journal} {\bibinfo  {journal} {Physics of Fluid}\ }\textbf
  {\bibinfo {volume} {31(12)}},\ \bibinfo {pages} {125112} (\bibinfo {year}
  {2019})}\BibitemShut {NoStop}%
\bibitem [{\citenamefont {Murata}, \citenamefont {Fukami},\ and\ \citenamefont
  {Fukagata}(2020)}]{Murataetal2020}%
  \BibitemOpen
  \bibfield  {author} {\bibinfo {author} {\bibfnamefont {T.}~\bibnamefont
  {Murata}}, \bibinfo {author} {\bibfnamefont {K.}~\bibnamefont {Fukami}},\
  and\ \bibinfo {author} {\bibfnamefont {K.}~\bibnamefont {Fukagata}},\
  }\bibfield  {title} {\enquote {\bibinfo {title} {Nonlinear mode decomposition
  with convolutional neural networks for fluid dynamics},}\ }\href@noop {}
  {\bibfield  {journal} {\bibinfo  {journal} {Journal of Fluid Mechanics}\
  }\textbf {\bibinfo {volume} {882}} (\bibinfo {year} {2020})}\BibitemShut
  {NoStop}%
\bibitem [{\citenamefont {Wang}\ \emph {et~al.}(2018)\citenamefont {Wang},
  \citenamefont {Xiao}, \citenamefont {Fang}, \citenamefont {Govindan},
  \citenamefont {Pain},\ and\ \citenamefont {Guo}}]{Wangetal2018}%
  \BibitemOpen
  \bibfield  {author} {\bibinfo {author} {\bibfnamefont {Z.}~\bibnamefont
  {Wang}}, \bibinfo {author} {\bibfnamefont {D.}~\bibnamefont {Xiao}}, \bibinfo
  {author} {\bibfnamefont {F.}~\bibnamefont {Fang}}, \bibinfo {author}
  {\bibfnamefont {R.}~\bibnamefont {Govindan}}, \bibinfo {author}
  {\bibfnamefont {C.~C.}\ \bibnamefont {Pain}},\ and\ \bibinfo {author}
  {\bibfnamefont {Y.}~\bibnamefont {Guo}},\ }\bibfield  {title} {\enquote
  {\bibinfo {title} {Model identification of reduced order fluid dynamics
  systems using deep learning},}\ }\href@noop {} {\bibfield  {journal}
  {\bibinfo  {journal} {International Journal for Numerical Methods in Fluids}\
  }\textbf {\bibinfo {volume} {86(4)}},\ \bibinfo {pages} {255--268} (\bibinfo
  {year} {2018})}\BibitemShut {NoStop}%
\bibitem [{\citenamefont {Mohan}\ and\ \citenamefont
  {Gaitonde}(2002)}]{Mohan&Gaitonde2002}%
  \BibitemOpen
  \bibfield  {author} {\bibinfo {author} {\bibfnamefont {A.~T.}\ \bibnamefont
  {Mohan}}\ and\ \bibinfo {author} {\bibfnamefont {D.~V.}\ \bibnamefont
  {Gaitonde}},\ }\bibfield  {title} {\enquote {\bibinfo {title} {A deep
  learning based approach to reduced order modeling for turbulent flow control
  using lstm neural networks},}\ }\href@noop {} {\bibfield  {journal} {\bibinfo
   {journal} {ArXiv Preprint ArXiv:1804.09269}\ } (\bibinfo {year}
  {2002})}\BibitemShut {NoStop}%
\bibitem [{\citenamefont {Mathey}\ \emph {et~al.}(2006)\citenamefont {Mathey},
  \citenamefont {Cokljat}, \citenamefont {Bertoglio},\ and\ \citenamefont
  {Sergent}}]{Matheyetal2006}%
  \BibitemOpen
  \bibfield  {author} {\bibinfo {author} {\bibfnamefont {F.}~\bibnamefont
  {Mathey}}, \bibinfo {author} {\bibfnamefont {D.}~\bibnamefont {Cokljat}},
  \bibinfo {author} {\bibfnamefont {J.~P.}\ \bibnamefont {Bertoglio}},\ and\
  \bibinfo {author} {\bibfnamefont {E.}~\bibnamefont {Sergent}},\ }\bibfield
  {title} {\enquote {\bibinfo {title} {Assessment of the vortex method for
  large eddy simulation inlet conditions},}\ }\href@noop {} {\bibfield
  {journal} {\bibinfo  {journal} {Progress in Computational Fluid Dynamics, an
  International Journal}\ }\textbf {\bibinfo {volume} {6(1–3)}},\ \bibinfo
  {pages} {58--67} (\bibinfo {year} {2006})}\BibitemShut {NoStop}%
\bibitem [{\citenamefont {Sergent}(2002)}]{Sergent2002}%
  \BibitemOpen
  \bibfield  {author} {\bibinfo {author} {\bibfnamefont {E.}~\bibnamefont
  {Sergent}},\ }\bibfield  {title} {\enquote {\bibinfo {title} {Vers une
  méthodologie de couplage entre la simulation des grandes échelles et les
  modèles statistiques [these de doctorat]},}\ }\href@noop {} {\bibfield
  {journal} {\bibinfo  {journal} {Ecully, Ecole centrale de Lyon}\ } (\bibinfo
  {year} {2002})}\BibitemShut {NoStop}%
\bibitem [{\citenamefont {Kolmogorov}(1941)}]{Kolmogorov1941}%
  \BibitemOpen
  \bibfield  {author} {\bibinfo {author} {\bibfnamefont {A.~N.}\ \bibnamefont
  {Kolmogorov}},\ }\bibfield  {title} {\enquote {\bibinfo {title} {The local
  structure of turbulence in incompressible viscous fluid for very large
  reynolds numbers},}\ }\href@noop {} {\bibfield  {journal} {\bibinfo
  {journal} {C. R. Acad. Sci. URSS}\ }\textbf {\bibinfo {volume} {30}},\
  \bibinfo {pages} {301--305} (\bibinfo {year} {1941})}\BibitemShut {NoStop}%
\bibitem [{\citenamefont {Moin}\ and\ \citenamefont
  {Mahesh}(1998)}]{Moin&Mahesh1998}%
  \BibitemOpen
  \bibfield  {author} {\bibinfo {author} {\bibfnamefont {P.}~\bibnamefont
  {Moin}}\ and\ \bibinfo {author} {\bibfnamefont {K.}~\bibnamefont {Mahesh}},\
  }\bibfield  {title} {\enquote {\bibinfo {title} {Direct numerical simulation:
  A tool in turbulence research},}\ }\href@noop {} {\bibfield  {journal}
  {\bibinfo  {journal} {Annual Review of Fluid Mechanics}\ }\textbf {\bibinfo
  {volume} {30(1)}},\ \bibinfo {pages} {539--578} (\bibinfo {year}
  {1998})}\BibitemShut {NoStop}%
\bibitem [{\citenamefont {Wilcox}(1993)}]{Wilcox1993}%
  \BibitemOpen
  \bibfield  {author} {\bibinfo {author} {\bibfnamefont {D.~C.}\ \bibnamefont
  {Wilcox}},\ }\bibfield  {title} {\enquote {\bibinfo {title} {Turbulence
  modeling for cfd},}\ }\href@noop {} {\bibfield  {journal} {\bibinfo
  {journal} {DCW Industries, Inc.}\ } (\bibinfo {year} {1993})}\BibitemShut
  {NoStop}%
\bibitem [{\citenamefont {Smagorinsky}(1987)}]{Smagorinsky1963}%
  \BibitemOpen
  \bibfield  {author} {\bibinfo {author} {\bibfnamefont {J.}~\bibnamefont
  {Smagorinsky}},\ }\bibfield  {title} {\enquote {\bibinfo {title} {General
  circulation experiments with the primitive equationsi. the basic
  experiment.}}\ }\href@noop {} {\bibfield  {journal} {\bibinfo  {journal}
  {Monthly Weather Review}\ }\textbf {\bibinfo {volume} {91(3)}},\ \bibinfo
  {pages} {99--164} (\bibinfo {year} {1987})}\BibitemShut {NoStop}%
\bibitem [{\citenamefont {Stolz}, \citenamefont {Adams},\ and\ \citenamefont
  {Kleiser}(2001)}]{Stolzetal2001}%
  \BibitemOpen
  \bibfield  {author} {\bibinfo {author} {\bibfnamefont {S.}~\bibnamefont
  {Stolz}}, \bibinfo {author} {\bibfnamefont {N.~A.}\ \bibnamefont {Adams}},\
  and\ \bibinfo {author} {\bibfnamefont {L.}~\bibnamefont {Kleiser}},\
  }\bibfield  {title} {\enquote {\bibinfo {title} {An approximate deconvolution
  model for large-eddy simulation with application to incompressible
  wall-bounded flows},}\ }\href@noop {} {\bibfield  {journal} {\bibinfo
  {journal} {Physics of Fluids}\ }\textbf {\bibinfo {volume} {13(4)}},\
  \bibinfo {pages} {997--1015} (\bibinfo {year} {2001})}\BibitemShut {NoStop}%
\bibitem [{\citenamefont {Wang}, \citenamefont {Thompson},\ and\ \citenamefont
  {Hu}(2019)}]{Wangetal2019}%
  \BibitemOpen
  \bibfield  {author} {\bibinfo {author} {\bibfnamefont {Y.}~\bibnamefont
  {Wang}}, \bibinfo {author} {\bibfnamefont {D.}~\bibnamefont {Thompson}},\
  and\ \bibinfo {author} {\bibfnamefont {Z.}~\bibnamefont {Hu}},\ }\bibfield
  {title} {\enquote {\bibinfo {title} {Effect of wall proximity on the flow
  over a cube and the implications for the noise emitted},}\ }\href@noop {}
  {\bibfield  {journal} {\bibinfo  {journal} {Physics of Fluids}\ }\textbf
  {\bibinfo {volume} {31(7)}},\ \bibinfo {pages} {077101} (\bibinfo {year}
  {2019})}\BibitemShut {NoStop}%
\bibitem [{\citenamefont {Spalart}\ \emph {et~al.}(1997)\citenamefont
  {Spalart}, \citenamefont {Jou}, \citenamefont {Strelets},\ and\ \citenamefont
  {Allmaras}}]{Spalartetal1997}%
  \BibitemOpen
  \bibfield  {author} {\bibinfo {author} {\bibfnamefont {P.}~\bibnamefont
  {Spalart}}, \bibinfo {author} {\bibfnamefont {W.}~\bibnamefont {Jou}},
  \bibinfo {author} {\bibfnamefont {M.}~\bibnamefont {Strelets}},\ and\
  \bibinfo {author} {\bibfnamefont {S.~R.}\ \bibnamefont {Allmaras}},\
  }\bibfield  {title} {\enquote {\bibinfo {title} {Comments on the feasibility
  of les for wings, and on a hybrid rans/les approach.}}\ }\href@noop {}
  {\bibfield  {journal} {\bibinfo  {journal} {Proceedings of First AFOSR
  International Conference on DNS/LES}\ } (\bibinfo {year} {1997})}\BibitemShut
  {NoStop}%
\bibitem [{\citenamefont {Spalart}(1997)}]{Spalart2000}%
  \BibitemOpen
  \bibfield  {author} {\bibinfo {author} {\bibfnamefont {P.~R.}\ \bibnamefont
  {Spalart}},\ }\bibfield  {title} {\enquote {\bibinfo {title} {Strategies for
  turbulence modelling and simulations.}}\ }\href@noop {} {\bibfield  {journal}
  {\bibinfo  {journal} {International Journal of Heat and Fluid Flow}\ }\textbf
  {\bibinfo {volume} {21(3)}},\ \bibinfo {pages} {252--263} (\bibinfo {year}
  {1997})}\BibitemShut {NoStop}%
\bibitem [{\citenamefont {Travin}\ \emph {et~al.}(2000)\citenamefont {Travin},
  \citenamefont {Shur}, \citenamefont {Strelets},\ and\ \citenamefont
  {Spalart}}]{Travinetal2000}%
  \BibitemOpen
  \bibfield  {author} {\bibinfo {author} {\bibfnamefont {A.}~\bibnamefont
  {Travin}}, \bibinfo {author} {\bibfnamefont {M.}~\bibnamefont {Shur}},
  \bibinfo {author} {\bibfnamefont {M.}~\bibnamefont {Strelets}},\ and\
  \bibinfo {author} {\bibfnamefont {P.}~\bibnamefont {Spalart}},\ }\bibfield
  {title} {\enquote {\bibinfo {title} {Detached-eddy simulations past a
  circular cylinder},}\ }\href@noop {} {\bibfield  {journal} {\bibinfo
  {journal} {Flow, Turbulence and Combustion}\ }\textbf {\bibinfo {volume}
  {63(1)}},\ \bibinfo {pages} {293--313} (\bibinfo {year} {2000})}\BibitemShut
  {NoStop}%
\bibitem [{\citenamefont {Spalart}(2009)}]{Spalart2009}%
  \BibitemOpen
  \bibfield  {author} {\bibinfo {author} {\bibfnamefont {R.~P.}\ \bibnamefont
  {Spalart}},\ }\bibfield  {title} {\enquote {\bibinfo {title} {Detached-eddy
  simulation.}}\ }\href@noop {} {\bibfield  {journal} {\bibinfo  {journal}
  {Annual Review of Fluid Mechanics}\ }\textbf {\bibinfo {volume} {41(1)}},\
  \bibinfo {pages} {181--202} (\bibinfo {year} {2009})}\BibitemShut {NoStop}%
\bibitem [{\citenamefont {Spalart}\ \emph {et~al.}(2006)\citenamefont
  {Spalart}, \citenamefont {Deck}, \citenamefont {Shur}, \citenamefont
  {Squires}, \citenamefont {Strelets},\ and\ \citenamefont
  {Travin}}]{Spalartetal2006}%
  \BibitemOpen
  \bibfield  {author} {\bibinfo {author} {\bibfnamefont {P.}~\bibnamefont
  {Spalart}}, \bibinfo {author} {\bibfnamefont {S.}~\bibnamefont {Deck}},
  \bibinfo {author} {\bibfnamefont {M.~L.}\ \bibnamefont {Shur}}, \bibinfo
  {author} {\bibfnamefont {K.~D.}\ \bibnamefont {Squires}}, \bibinfo {author}
  {\bibfnamefont {M.~K.}\ \bibnamefont {Strelets}},\ and\ \bibinfo {author}
  {\bibfnamefont {A.}~\bibnamefont {Travin}},\ }\bibfield  {title} {\enquote
  {\bibinfo {title} {A new version of detached-eddy simulation, resistant to
  ambiguous grid densities.}}\ }\href@noop {} {\bibfield  {journal} {\bibinfo
  {journal} {Theoretical and Computational Fluid Dynamics}\ }\textbf {\bibinfo
  {volume} {20(3)}},\ \bibinfo {pages} {181} (\bibinfo {year}
  {2006})}\BibitemShut {NoStop}%
\bibitem [{\citenamefont {Shur}\ \emph {et~al.}(2008)\citenamefont {Shur},
  \citenamefont {Spalart}, \citenamefont {Strelets},\ and\ \citenamefont
  {Travin}}]{Shuretal2008}%
  \BibitemOpen
  \bibfield  {author} {\bibinfo {author} {\bibfnamefont {M.~L.}\ \bibnamefont
  {Shur}}, \bibinfo {author} {\bibfnamefont {P.~R.}\ \bibnamefont {Spalart}},
  \bibinfo {author} {\bibfnamefont {M.~K.}\ \bibnamefont {Strelets}},\ and\
  \bibinfo {author} {\bibfnamefont {A.~K.}\ \bibnamefont {Travin}},\ }\bibfield
   {title} {\enquote {\bibinfo {title} {A hybrid rans-les approach with
  delayed-des and wall-modelled les capabilities},}\ }\href@noop {} {\bibfield
  {journal} {\bibinfo  {journal} {International Journal of Heat and Fluid
  Flow}\ }\textbf {\bibinfo {volume} {29(6)}},\ \bibinfo {pages} {1638--1649}
  (\bibinfo {year} {2008})}\BibitemShut {NoStop}%
\bibitem [{\citenamefont {Berkooz}, \citenamefont {Holmes},\ and\ \citenamefont
  {Lumley}(1993)}]{Berkoozetal1993}%
  \BibitemOpen
  \bibfield  {author} {\bibinfo {author} {\bibfnamefont {G.}~\bibnamefont
  {Berkooz}}, \bibinfo {author} {\bibfnamefont {P.}~\bibnamefont {Holmes}},\
  and\ \bibinfo {author} {\bibfnamefont {J.~L.}\ \bibnamefont {Lumley}},\
  }\bibfield  {title} {\enquote {\bibinfo {title} {The proper orthogonal
  decomposition in the analysis of turbulent flows},}\ }\href@noop {}
  {\bibfield  {journal} {\bibinfo  {journal} {Annual Review of Fluid
  Mechanics}\ }\textbf {\bibinfo {volume} {25(1)}},\ \bibinfo {pages}
  {539--575} (\bibinfo {year} {1993})}\BibitemShut {NoStop}%
\bibitem [{\citenamefont {Sirovich}(1987)}]{Sirovich1987}%
  \BibitemOpen
  \bibfield  {author} {\bibinfo {author} {\bibfnamefont {L.}~\bibnamefont
  {Sirovich}},\ }\bibfield  {title} {\enquote {\bibinfo {title} {Turbulence and
  the dynamics of coherent structures. i. coherent structures},}\ }\href@noop
  {} {\bibfield  {journal} {\bibinfo  {journal} {Quarterly of Applied
  Mathematics}\ }\textbf {\bibinfo {volume} {45(3)}},\ \bibinfo {pages}
  {561--571} (\bibinfo {year} {1987})}\BibitemShut {NoStop}%
\bibitem [{\citenamefont {Hochreiter}\ and\ \citenamefont
  {Schmidhuber}(1997)}]{Hochreiter&Schmidhuber1997}%
  \BibitemOpen
  \bibfield  {author} {\bibinfo {author} {\bibfnamefont {S.}~\bibnamefont
  {Hochreiter}}\ and\ \bibinfo {author} {\bibfnamefont {J.}~\bibnamefont
  {Schmidhuber}},\ }\bibfield  {title} {\enquote {\bibinfo {title} {Long
  short-term memory},}\ }\href@noop {} {\bibfield  {journal} {\bibinfo
  {journal} {Neural Computation}\ }\textbf {\bibinfo {volume} {9(8)}},\
  \bibinfo {pages} {1735--1780} (\bibinfo {year} {1997})}\BibitemShut {NoStop}%
\bibitem [{\citenamefont {Rumelhart}, \citenamefont {Hinton},\ and\
  \citenamefont {Williams}(1986)}]{Rumelhartetal1986}%
  \BibitemOpen
  \bibfield  {author} {\bibinfo {author} {\bibfnamefont {D.~E.}\ \bibnamefont
  {Rumelhart}}, \bibinfo {author} {\bibfnamefont {G.~E.}\ \bibnamefont
  {Hinton}},\ and\ \bibinfo {author} {\bibfnamefont {R.~J.}\ \bibnamefont
  {Williams}},\ }\bibfield  {title} {\enquote {\bibinfo {title} {Learning
  representations by back-propagating errors},}\ }\href@noop {} {\bibfield
  {journal} {\bibinfo  {journal} {Nature}\ }\textbf {\bibinfo {volume}
  {323(6088)}},\ \bibinfo {pages} {533--536} (\bibinfo {year}
  {1986})}\BibitemShut {NoStop}%
\bibitem [{\citenamefont {Graves}(2012)}]{Graves2012}%
  \BibitemOpen
  \bibfield  {author} {\bibinfo {author} {\bibfnamefont {A.}~\bibnamefont
  {Graves}},\ }\bibfield  {title} {\enquote {\bibinfo {title} {Long short-term
  memory.}}\ }\href@noop {} {\bibfield  {journal} {\bibinfo  {journal} {In A.
  Graves (Ed.), Supervised Sequence Labelling with Recurrent Neural Networks}\
  ,\ \bibinfo {pages} {37--–45}} (\bibinfo {year} {2012})}\BibitemShut
  {NoStop}%
\bibitem [{\citenamefont {Graves}, \citenamefont {Fernández},\ and\
  \citenamefont {Schmidhuber}(2005)}]{Gravesetal2005}%
  \BibitemOpen
  \bibfield  {author} {\bibinfo {author} {\bibfnamefont {A.}~\bibnamefont
  {Graves}}, \bibinfo {author} {\bibfnamefont {S.}~\bibnamefont {Fernández}},\
  and\ \bibinfo {author} {\bibfnamefont {J.}~\bibnamefont {Schmidhuber}},\
  }\bibfield  {title} {\enquote {\bibinfo {title} {Bidirectional lstm networks
  for improved phoneme classification and recognition},}\ }\href@noop {}
  {\bibfield  {journal} {\bibinfo  {journal} {In W. Duch, J. Kacprzyk, E. Oja,
  \& S. Zadrożny (Eds.), Artificial Neural Networks: Formal Models and Their
  Applications – ICANN 2005}\ ,\ \bibinfo {pages} {799–--804}} (\bibinfo
  {year} {2005})}\BibitemShut {NoStop}%
\bibitem [{\citenamefont {Graves}\ and\ \citenamefont
  {Schmidhuber}(2005)}]{Graves&Schmidhuber2005}%
  \BibitemOpen
  \bibfield  {author} {\bibinfo {author} {\bibfnamefont {A.}~\bibnamefont
  {Graves}}\ and\ \bibinfo {author} {\bibfnamefont {J.}~\bibnamefont
  {Schmidhuber}},\ }\bibfield  {title} {\enquote {\bibinfo {title} {Framewise
  phoneme classification with bidirectional lstm and other neural network
  architectures},}\ }\href@noop {} {\bibfield  {journal} {\bibinfo  {journal}
  {Neural Networks}\ }\textbf {\bibinfo {volume} {18(5)}},\ \bibinfo {pages}
  {602--610} (\bibinfo {year} {2005})}\BibitemShut {NoStop}%
\bibitem [{\citenamefont {Huang}, \citenamefont {Xu},\ and\ \citenamefont
  {Yu}(2015)}]{Huangetal2015}%
  \BibitemOpen
  \bibfield  {author} {\bibinfo {author} {\bibfnamefont {Z.}~\bibnamefont
  {Huang}}, \bibinfo {author} {\bibfnamefont {W.}~\bibnamefont {Xu}},\ and\
  \bibinfo {author} {\bibfnamefont {K.}~\bibnamefont {Yu}},\ }\bibfield
  {title} {\enquote {\bibinfo {title} {Bidirectional lstm-crf models for
  sequence tagging},}\ }\href@noop {} {\bibfield  {journal} {\bibinfo
  {journal} {ArXiv:1508.01991 [Cs]}\ } (\bibinfo {year} {2015})}\BibitemShut
  {NoStop}%
\bibitem [{\citenamefont {Marchi}\ \emph {et~al.}(2014)\citenamefont {Marchi},
  \citenamefont {Ferroni}, \citenamefont {Eyben}, \citenamefont {Gabrielli},
  \citenamefont {Squartini},\ and\ \citenamefont {Schuller}}]{Marchietal2014}%
  \BibitemOpen
  \bibfield  {author} {\bibinfo {author} {\bibfnamefont {E.}~\bibnamefont
  {Marchi}}, \bibinfo {author} {\bibfnamefont {G.}~\bibnamefont {Ferroni}},
  \bibinfo {author} {\bibfnamefont {F.}~\bibnamefont {Eyben}}, \bibinfo
  {author} {\bibfnamefont {L.}~\bibnamefont {Gabrielli}}, \bibinfo {author}
  {\bibfnamefont {S.}~\bibnamefont {Squartini}},\ and\ \bibinfo {author}
  {\bibfnamefont {B.}~\bibnamefont {Schuller}},\ }\bibfield  {title} {\enquote
  {\bibinfo {title} {Multi-resolution linear prediction based features for
  audio onset detection with bidirectional lstm neural networks},}\ }\href@noop
  {} {\bibfield  {journal} {\bibinfo  {journal} {2014 IEEE International
  Conference on Acoustics, Speech and Signal Processing (ICASSP)}\ ,\ \bibinfo
  {pages} {2164--2168}} (\bibinfo {year} {2014})}\BibitemShut {NoStop}%
\bibitem [{\citenamefont {Weiss}, \citenamefont {Khoshgoftaar},\ and\
  \citenamefont {Wang}(2016)}]{Weissetal2016}%
  \BibitemOpen
  \bibfield  {author} {\bibinfo {author} {\bibfnamefont {K.}~\bibnamefont
  {Weiss}}, \bibinfo {author} {\bibfnamefont {T.~M.}\ \bibnamefont
  {Khoshgoftaar}},\ and\ \bibinfo {author} {\bibfnamefont {D.}~\bibnamefont
  {Wang}},\ }\bibfield  {title} {\enquote {\bibinfo {title} {A survey of
  transfer learning},}\ }\href@noop {} {\bibfield  {journal} {\bibinfo
  {journal} {Journal of Big Data}\ }\textbf {\bibinfo {volume} {3(1)}},\
  \bibinfo {pages} {9} (\bibinfo {year} {2016})}\BibitemShut {NoStop}%
\bibitem [{\citenamefont {Kingma}\ and\ \citenamefont
  {Ba}(2014)}]{Kingma&Ba2014}%
  \BibitemOpen
  \bibfield  {author} {\bibinfo {author} {\bibfnamefont {B.~P.}\ \bibnamefont
  {Kingma}}\ and\ \bibinfo {author} {\bibfnamefont {J.}~\bibnamefont {Ba}},\
  }\bibfield  {title} {\enquote {\bibinfo {title} {Adam: A method for
  stochastic optimization.}}\ }\href@noop {} {\bibfield  {journal} {\bibinfo
  {journal} {https://arxiv.org/abs/1412.6980v9}\ } (\bibinfo {year}
  {2014})}\BibitemShut {NoStop}%
\bibitem [{\citenamefont {Cavers}(2017)}]{Cavers2017}%
  \BibitemOpen
  \bibfield  {author} {\bibinfo {author} {\bibfnamefont {J.}~\bibnamefont
  {Cavers}},\ }\bibfield  {title} {\enquote {\bibinfo {title} {Chaos and time
  series analysis: Optimization of the poincaré section and distinguishing
  between deterministic and stochastic time series},}\ }\href@noop {}
  {\bibfield  {journal} {\bibinfo  {journal} {Electronic Thesis and
  Dissertations.}\ } (\bibinfo {year} {2017})}\BibitemShut {NoStop}%
\bibitem [{\citenamefont {Wolf}\ \emph {et~al.}(1985)\citenamefont {Wolf},
  \citenamefont {Swift}, \citenamefont {Swinney},\ and\ \citenamefont
  {Vastano}}]{Wolfetal1985}%
  \BibitemOpen
  \bibfield  {author} {\bibinfo {author} {\bibfnamefont {A.}~\bibnamefont
  {Wolf}}, \bibinfo {author} {\bibfnamefont {J.~B.}\ \bibnamefont {Swift}},
  \bibinfo {author} {\bibfnamefont {H.}~\bibnamefont {Swinney}},\ and\ \bibinfo
  {author} {\bibfnamefont {J.~A.}\ \bibnamefont {Vastano}},\ }\bibfield
  {title} {\enquote {\bibinfo {title} {Determining lyapunov exponents from a
  time series},}\ }\href@noop {} {\bibfield  {journal} {\bibinfo  {journal}
  {Physica D: Nonlinear Phenomena}\ }\textbf {\bibinfo {volume} {16(3)}},\
  \bibinfo {pages} {285--317} (\bibinfo {year} {1985})}\BibitemShut {NoStop}%
\bibitem [{\citenamefont {Franke}\ \emph {et~al.}(2004)\citenamefont {Franke},
  \citenamefont {Hirsch}, \citenamefont {Jensen}, \citenamefont {Krüs},
  \citenamefont {Schatzmann}, \citenamefont {Westbury}, \citenamefont {Miles},
  \citenamefont {Wisse},\ and\ \citenamefont {Wright}}]{Frankeetal2004}%
  \BibitemOpen
  \bibfield  {author} {\bibinfo {author} {\bibfnamefont {J.}~\bibnamefont
  {Franke}}, \bibinfo {author} {\bibfnamefont {C.}~\bibnamefont {Hirsch}},
  \bibinfo {author} {\bibfnamefont {A.}~\bibnamefont {Jensen}}, \bibinfo
  {author} {\bibfnamefont {H.}~\bibnamefont {Krüs}}, \bibinfo {author}
  {\bibfnamefont {M.}~\bibnamefont {Schatzmann}}, \bibinfo {author}
  {\bibfnamefont {P.}~\bibnamefont {Westbury}}, \bibinfo {author}
  {\bibfnamefont {S.}~\bibnamefont {Miles}}, \bibinfo {author} {\bibfnamefont
  {J.}~\bibnamefont {Wisse}},\ and\ \bibinfo {author} {\bibfnamefont {N.~G.}\
  \bibnamefont {Wright}},\ }\bibfield  {title} {\enquote {\bibinfo {title}
  {Recommendations on the use of cfd in wind engineering, cost action c14},}\
  }\href@noop {} {\bibfield  {journal} {\bibinfo  {journal} {European Science
  Foundation COST Office}\ }\textbf {\bibinfo {volume} {14}},\ \bibinfo {pages}
  {c1} (\bibinfo {year} {2004})}\BibitemShut {NoStop}%
\bibitem [{\citenamefont {Saha}(2013)}]{Saha2013}%
  \BibitemOpen
  \bibfield  {author} {\bibinfo {author} {\bibfnamefont {A.~K.}\ \bibnamefont
  {Saha}},\ }\bibfield  {title} {\enquote {\bibinfo {title} {Unsteady flow past
  a finite square cylinder mounted on a wall at low reynolds number},}\
  }\href@noop {} {\bibfield  {journal} {\bibinfo  {journal} {Computers \&
  Fluids}\ }\textbf {\bibinfo {volume} {88}},\ \bibinfo {pages} {599--615}
  (\bibinfo {year} {2013})}\BibitemShut {NoStop}%
\bibitem [{\citenamefont {Schmitt}(2007)}]{Schmitt2007}%
  \BibitemOpen
  \bibfield  {author} {\bibinfo {author} {\bibfnamefont {F.~G.}\ \bibnamefont
  {Schmitt}},\ }\bibfield  {title} {\enquote {\bibinfo {title} {About
  boussinesq's turbulent viscosity hypothesis: Historical remarks and a direct
  evaluation of its validity},}\ }\href@noop {} {\bibfield  {journal} {\bibinfo
   {journal} {Comptes Rendus Mécanique}\ }\textbf {\bibinfo {volume}
  {335(9)}},\ \bibinfo {pages} {617--627} (\bibinfo {year} {2007})}\BibitemShut
  {NoStop}%
\bibitem [{\citenamefont {Weiss}(2019)}]{Weiss2019}%
  \BibitemOpen
  \bibfield  {author} {\bibinfo {author} {\bibfnamefont {J.}~\bibnamefont
  {Weiss}},\ }\bibfield  {title} {\enquote {\bibinfo {title} {A tutorial on the
  proper orthogonal decomposition},}\ }\href@noop {} {\bibfield  {journal}
  {\bibinfo  {journal} {In AIAA Aviation 2019 Forum (Vol. 1–0). American
  Institute of Aeronautics and Astronautics}\ } (\bibinfo {year}
  {2019})}\BibitemShut {NoStop}%
\end{thebibliography}%

\end{document}